\documentclass[final,3p,times,twocolumn]{elsarticle}
\usepackage{graphicx}
\usepackage{amssymb}
\usepackage{amsmath}
\usepackage{mathtools}
\usepackage{color, colortbl}
\usepackage[squaren,Gray]{SIunits}
\usepackage{subfigure}
\usepackage{placeins}
\usepackage{multirow}
\biboptions{comma,square,sort&compress}

\newcounter{bla}

\journal{Computer Physics Communications}

\begin{document}

\definecolor{Gray}{gray}{0.9}

\begin{frontmatter}

\title{EPW: Electron-phonon coupling, transport and superconducting properties using maximally localized Wannier functions}

\author[oxford]{S. Ponc\'e \corref{cor1}}
\ead{samuel.pon@gmail.com}
\author[binghamton]{E. R. Margine }
\author[oxford]{C. Verdi}
\author[oxford]{F. Giustino\corref{cor1}}
\ead{feliciano.giustino@materials.ox.ac.uk}

\address[oxford]{Department of Materials, University of Oxford, Parks Road, Oxford, OX1 3PH, UK}
\address[binghamton]{Department of Physics, Applied Physics and Astronomy, Binghamton University-SUNY, Binghamton, New York 13902, USA}

\cortext[cor1]{Corresponding authors}

\begin{abstract}
The \texttt{EPW} (\underline{E}lectron-\underline{P}honon coupling using \underline{W}annier functions) software is a \texttt{Fortran90} code that uses density-functional perturbation theory and maximally localized Wannier functions for computing
electron-phonon couplings and related properties in solids accurately and efficiently.
The \texttt{EPW} v4 program can be used to compute electron and phonon self-energies, linewidths, electron-phonon scattering rates, electron-phonon coupling strengths, transport spectral functions, electronic velocities, resistivity, anisotropic superconducting gaps and spectral functions within the Migdal-Eliashberg theory.
The code now supports spin-orbit coupling, time-reversal symmetry in non-centrosymmetric crystals, polar materials, and $\mathbf{k}$ and $\mathbf{q}$-point parallelization.
Considerable effort was dedicated to optimization and parallelization, achieving almost a ten times speedup with respect to previous releases.
A computer test farm was implemented to ensure stability and portability of the code on the most popular compilers and architectures. 
Since April 2016, version 4 of the \texttt{EPW} code is fully integrated in and distributed with the \texttt{Quantum ESPRESSO} package, and can be downloaded through QE-forge at \texttt{http://qe-forge.org/gf/project/q-e}. 
\end{abstract}


\end{frontmatter}



{\bf NEW VERSION PROGRAM SUMMARY}

\begin{small}
\noindent
{\em Manuscript Title: EPW: Electron-phonon coupling, transport and superconducting properties using maximally localized Wannier functions}                                       \\
{\em Authors: S. Ponc\'e, E. R. Margine, C. Verdi and F. Giustino}                         \\
{\em Program Title: EPW}                                          \\
{\em Journal Reference:}                                      \\
{\em Catalogue identifier:}                                   \\
{\em Licensing provisions: GNU General Public License 3}                                   \\
{\em Programming language: Fortran 90, MPI}                          \\
{\em Computer: Non-specific}                                               \\
{\em Operating system: Unix/Linux}                                       \\
{\em RAM: Typically 2GB/core}                                               \\
{\em Number of processors used: Typically between 16 and 512}                              
{\em Classification: 7.3, 7.8, 7.9}                                         \\
{\em External routines/libraries: LAPACK, BLAS, MPI, FFTW, Quantum-ESPRESSO package [1]}   \\
{\em Catalogue identifier of previous version: AEHA\_v1\_0 }             \\
{\em Journal reference of previous version: Computer Physics Communications 181 (2010) 2140-2148 }                  \\
{\em Does the new version supersede the previous version?: Yes}   \\
{\em Nature of problem: Calculation of electron and phonon self-energies, linewidths, electron-phonon scattering rates, electron-phonon coupling strengths, transport spectral functions, electronic velocities, resistivity, anisotropic superconducting gaps and spectral functions within the Migdal-Eliashberg theory. }\\
   \\
{\em Solution method: The code relies on density-functional perturbation theory and maximally localized Wannier functions}\\
   \\
{\em Reasons for the new version: New features (listed in the paper) and optimization of the code}\\
   \\
{\em Summary of revisions: Recent developments and new functionalities are described in Section~\ref{Functionalities} of the paper}\\
   \\
{\em Running time: Up to several hours on several tens of processors}\\
   \\

\end{small}

\section{Introduction}
\label{Introduction}

The electron-phonon coupling plays a crucial role in a wealth of phenomena, including the electron velocity renormalization~\cite{Park2007} and lifetimes ~\cite{Park2009,Eiguren2003}, phonon softening~\cite{Pisana2007} and lifetimes~\cite{Liao2015,Lazzeri2006}, phonon-assisted absorption~\cite{Kioupakis2010,Noffsinger2012,Zacharias2015}, critical temperature in conventional superconductors~\cite{Choi2002,Luders2005,Marques2005,Margine2014,Margine2016},  resistivity~\cite{Restrepo2009,Park2014} and the renormalization of electronic excitation energies~\cite{Franceschetti2007,Kamisaka2008,Ibrahim2008,Ramirez2006,Ramirez2008,Capaz2005,Patrick2013,Han2013,
Monserrat2013,Monserrat2014a,Antonius2014,Monserrat2016,Marini2008,Giustino2010,Gonze2011,Cannuccia2011,Cannuccia2012,Kawai2014,Ponce2014,Ponce2014a,Marini2015,Ponce2015,Antonius2015}. 

Nowadays, most first-principles software packages for periodic solids provide ground-state electronic properties through density-functional theory (DFT)~\cite{Hohenberg1964,Kohn1965}. Vibrational properties such as phonon frequencies and electron-phonon matrix elements are usually computed within density-functional perturbation theory (DFPT)~\cite{Baroni1987,Savrasov1992,Gonze1997,Baroni2001}.
All the aforementioned properties require very fine sampling of the electron and phonon wavevectors to achieve numerical convergence of the required Brillouin Zone (BZ) integrals~\cite{Giustino2007,Ponce2014,Ponce2015}. 

There has been a renewed interest in developing new methods and software to address this problem in a computationally feasible way. 
Electron-phonon matrix elements computed using DFPT can be linearly interpolated~\cite{Li2015} or extrapolated~\cite{Ponce2015}. The use of supercell with localized basis set~\cite{Gunst2016} and Fermi surface harmonics~\cite{Allen1976a,Eiguren2014} have also been investigated.   

The \texttt{EPW} code addresses the problem of computing electron-phonon matrix elements on a fine sampling of the BZ through maximally localized Wannier functions (MLWFs)~\cite{Marzari1997, Souza2001, Mostofi2008} and generalized Fourier interpolation~\cite{Giustino2007}. 
After the development of the \texttt{EPW} software similar efforts also based on the MLWFs of electron-phonon matrix elements were reported~\cite{Calandra2010,Sjakste2015,Calandra2015}.

The \texttt{EPW} project~\cite{EPWwebsite} started in 2006 with a first public release of the code in 2010 under a GNU General Public License (GPL)~\cite{GNU}. Most of the theoretical development of the code is described in Refs.~\cite{Giustino2007,Margine2013,Verdi2015},
while the \texttt{EPW} code and basic functionalities have been described Ref.~\cite{Noffsinger2010}. In this paper we will therefore focus on the new developments and functionalities added since the publication of Ref.~\cite{Noffsinger2010}. 

The manuscript is organized as follows. Section~\ref{Functionalities} summarizes the capabilities and new features of the code. Section~\ref{theory} outlines the theory and main equations used to compute standard electron-phonon properties, while the case of polar materials is discussed in Section~\ref{polarmat}.
The implementation of the superconducting formalism based on the Migdal-Eliashberg theory is presented in Section~\ref{supercond}.   
Basic electronic transport properties and associated electronic velocities are described in Sections~\ref{transportProp} and \ref{veloelectro}.
The new developments related to parallelization and speedup of the \texttt{EPW} software are presented in Section~\ref{para}.
The code is currently tested on a wide range of compilers and architectures using a Buildbot test farm, as described in Section~\ref{testfarm}.
Finally, we highlight in Section~\ref{examplessection} the new capabilities of \texttt{EPW} through five physically relevant examples: spectral functions and linewidths of B-doped diamond; scattering rate of undoped Si; spectral function and electronic resistivity of Pb with and without spin-orbit coupling; electron-phonon matrix elements for the polar semiconductor GaN; and superconducting properties of MgB$_2$ based on the Migdal-Eliashberg theory.

\section{Functionalities and technical release}
\label{Functionalities}

The \texttt{EPW} software is a freely available \texttt{Fortran90} code for periodic systems that relies on DFPT and MLWFs to compute properties related to the electron-phonon coupling on very fine electron (\textbf{k}) and phonon (\textbf{q}) wavevector grids. 

The software is now fully integrated into the \texttt{Quantum ESPRESSO}~\cite{Giannozzi2009} software package, and uses the capabilities of the \texttt{wannier90}~\cite{Mostofi2008} code as an internal library. 
\texttt{EPW} supports norm-conserving pseudopotentials with or without non-linear core-corrections~\cite{Perdew1981,Troullier1991,Fuchs1999}
as well as Hamann multi-projector norm-conserving potentials~\cite{Bachelet1982,Hamann2013}.
Many different exchange-correlation functionals are supported in the framework of the local-density (LDA)~\cite{Ceperley1980,Perdew1981} or generalized-gradient approximation (GGA)~\cite{Perdew1992a}.

The \texttt{EPW} code is based on the same file format as \texttt{Quantum ESPRESSO} and therefore supports text, binaries and Extensible Markup Language (XML).

The current version 4.0 of \texttt{EPW} includes around 16800 lines of code that rely on \texttt{BLAS}, \texttt{LAPACK} and \texttt{FFTW3} external libraries, as well as on several routines of \texttt{Quantum ESPRESSO}. The code is parallelized using a message passing interface (\texttt{MPI}) library.
The software has its own dedicated website \texttt{http://epw.org.uk} and forum \texttt{http://epwforum.uk}. 

\texttt{EPW} can calculate:
\begin{itemize}
	\item the electron linewidth $\Sigma_{n\mathbf{k}}''$ and phonon linewidth $\Pi_{\mathbf{q}\nu}''$, see Eqs.~\eqref{elselfenergy} and \eqref{phselfenergy};	
	\item the electron scattering rate $\tau_{n\mathbf{k}}^{-1}$, see Eq.~\eqref{scatteringRate};
    \item the nesting function $\zeta_{\mathbf{q}}$, see Eq.~\eqref{nestingfct};
	\item the electron-phonon coupling strength $\lambda_{\mathbf{q}\nu}$, see Eq.~\eqref{lambdaEQ}.
\end{itemize}

Recent developments and new functionalities since \texttt{EPW} v. 2.3 include:
\begin{itemize}	
	\item the code is integrated in the latest \texttt{Quantum ESPRESSO} distribution (v. 5.3);
	\item the spin-orbit coupling is now implemented;
	\item the treatment of time-reversal symmetry of non-centrosymmetric crystal is now supported;
	\item the polar electron-phonon vertex interpolated using MLWFs now correctly captures the long-wavelength singularity in polar materials. This enables calculations for semi-conductors and insulators in addition to metals;
\end{itemize}
\begin{itemize}
	\item the calculation of the Allen-Dynes superconductivity critical temperature, see Eq.~\eqref{ADTc};
	\item the anisotropic Eliashberg spectral function $\alpha^2F$, see Eq.~\eqref{Elishberga2F};	
	\item the anisotropic superconducting gap $\Delta_{n\mathbf{k}}$ within the Eliashberg theory, see Eqs.~\eqref{coupledAnisoEqs1} and \eqref{coupledAnisoEqs2};
	\item the specific heat in the superconducting state, see Eq.~\eqref{specific_heat};
	\item the tunneling density of states, see Eq.~\eqref{tunnelingdensity};
	\item Ziman's resistivity formula, see Eq.~\eqref{resistivitytr};			
	\item the transport spectral function $\alpha_{\text{tr}}^2 F$, see Eq.~\eqref{eliashtransport};	
	\item electronic velocities in the \textit{local approximation} $\tilde{v}_{mn\mathbf{k}}$, see Eq.~\eqref{finalvelo};
	\item tutorials, documentation and website have been revamped to improve user experience. A forum has been created;
	\item a test farm has been set up in order to ensure portability of the code on many architectures and compilers.	
\end{itemize}

\begin{figure}
 \centering
    \includegraphics[width=0.3\textwidth]{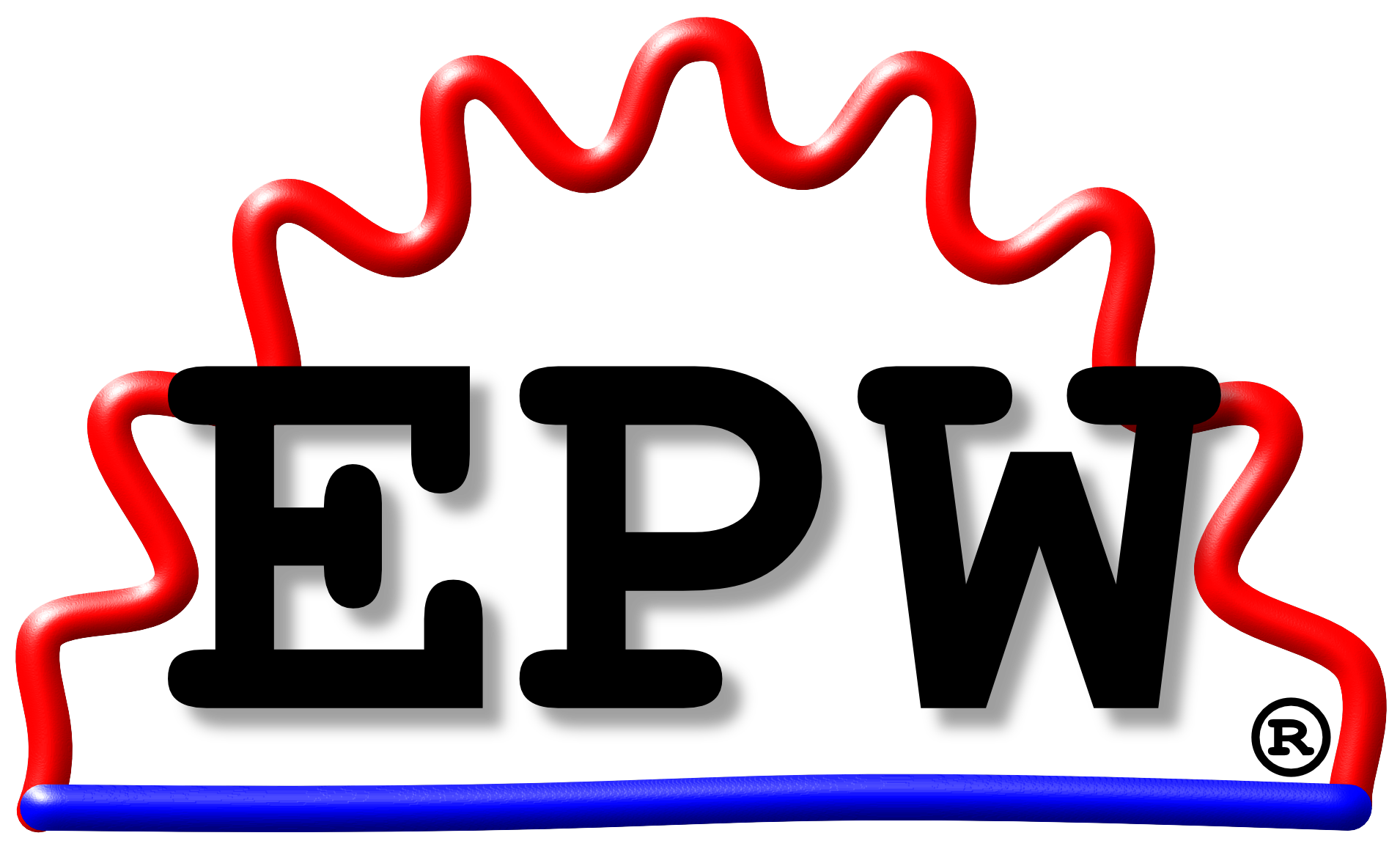}
  \caption[]{\label{fig:trademarkEPW} The \texttt{EPW} name and logo are now an officially registered trademark. The software is free and under a GNU license.  }
\end{figure}

The \texttt{EPW} name and logo (see Fig.~\ref{fig:trademarkEPW}) are now an officially registered trademark with number UK00003123373. The software is nonetheless free and released under a GNU General Public License. Companies or individuals who wish to use the \texttt{EPW} name and/or logo should contact Prof. F. Giustino (\texttt{feliciano.giustino@materials.ox.ac.uk}).

\section{Electron and phonon self-energies and electron-phonon coupling strength}\label{theory}

In this section we give a brief summary of the basic physical quantities that can be computed using the \texttt{EPW} software. Detailed derivation of the equations can be found in Refs.~\cite{Giustino2007} and \cite{Margine2013}. All equations presented in this work are expressed in atomic unit where $m_e=\hbar=k_{\rm B}=1$.

The first-order electron-phonon matrix element can be computed using DFPT as:	
\begin{equation}\label{elphmatrixoriginal}
	g_{mn,\nu}(\mathbf{k,k'}) =  \frac{1}{\sqrt{2 \omega_{\mathbf{k'-k}\nu} }} \langle \psi_{m\mathbf{k}'} | \partial_{\mathbf{k'-k}\nu}V | \psi_{n\mathbf{k}}\rangle,
\end{equation}
which quantifies a scattering process between the Khon-Sham states $m\bf{k'}$ and $n\bf{k}$. The matrix elements can conveniently be written as $g_{mn,\nu}(\mathbf{k,q})$ instead of $g_{mn,\nu}(\mathbf{k,k+q})$ using the phonon wavevector $\bf{q}={\bf{k'-k}}$: 					
\begin{equation}\label{elphmatrix}
	g_{mn,\nu}(\mathbf{k,q}) =  \frac{1}{\sqrt{2 \omega_{\mathbf{q}\nu}} } \langle \psi_{m\mathbf{k+q}}  | \partial_{\mathbf{q}\nu}V | \psi_{n\mathbf{k}} \rangle,
\end{equation}
where $\partial_{{\bf q}\nu}V$ is the derivative of the self-consistent potential associated with a phonon of wavevector \textbf{q}, branch index $\nu$, and frequency $\omega_{\mathbf{q}\nu}$. 
The nucleus masses are already included in the phonon eigenmodes.
The $\psi_{n\mathbf{k}}$ is the electronic wavefunction for band $n$ and wavevector $\mathbf{k}$, with $\varepsilon_{n\mathbf{k}}$ being the associated eigenvalue.     
The electron ($\Sigma$) and phonon ($\Pi$) self-energies at the temperature $T$ for metals and doped semiconductors are given by~\cite{Noffsinger2010,Giustino2016}:
\begin{multline}\label{elselfenergy}
   \Sigma_{n\mathbf{k}}(\omega,T) =  \sum_{m\nu} \int_{\rm BZ} \frac{d\mathbf{q}}{\Omega_{\rm BZ}} | g_{mn,\nu}(\mathbf{k,q}) |^2 \\
    \times \bigg[\frac{n_{\mathbf{q}\nu}(T)+f_{m\mathbf{k+q}}(T)}{\omega - (\varepsilon_{m\mathbf{k+q}}-\varepsilon_{\rm F}) +\omega_{\mathbf{q}\nu}+i\delta}\\
    + \frac{n_{\mathbf{q}\nu}(T)+1-f_{m\mathbf{k+q}}(T)}{\omega - (\varepsilon_{m\mathbf{k+q}}-\varepsilon_{\rm F}) - \omega_{\mathbf{q}\nu}+i\delta}\bigg], 
\end{multline}      
\begin{multline}\label{phselfenergy}
  \Pi_{\mathbf{q}\nu}(\omega,T) =  2\sum_{mn} \int_{\rm BZ} \frac{d\mathbf{k}}{\Omega_{\rm BZ}} | g_{mn,\nu}(\mathbf{k,q}) |^2 \\
    \times \frac{f_{n\mathbf{k}}(T)-f_{m\mathbf{k+q}}(T)}{\varepsilon_{n\mathbf{k}} - \varepsilon_{m\mathbf{k+q}}-\omega-i\delta}, 
\end{multline}  
where the factor $2$ accounts for the spin degeneracy, $\varepsilon_{\rm F}$ is the Fermi energy, $n_{\mathbf{q}\nu}(T)=[\exp(\frac{\omega_{\mathbf{q}\nu}}{T})-1]^{-1}$ is the Bose-Einstein distribution, $f_{n\mathbf{k}}(T)$ is the electronic occupation at wavevector $\mathbf{k}$ and band $n$ and, $\delta$ is a small positive real parameter which guarantees the correct analytical structure of the self-energies, and avoids numerical instabilities. The physical interpretation of $\delta$ is related to the finite lifetime of the electronic states. The integrals extend over the Brillouin Zone (BZ) of volume $\Omega_{\rm BZ}$.

Note that Eq.~\eqref{elselfenergy} does not contain the real and static Debye-Waller term that is usually found when studying semiconductors or insulators~\cite{Allen1976,Allen1981,Allen1983,Ponce2014a,Giustino2016}. 
Instead, when studying metals and doped semiconductors, we compute the real part of the electron self energy as: 
\begin{equation}\label{realpartselfenergy}
 \tilde{\Sigma}_{n\mathbf{k}}'(\omega,T) = \Sigma'_{n\mathbf{k}}(\omega,T) - \Sigma_{n\mathbf{k}}'(\omega=\varepsilon_{\rm F},T),
\end{equation}
where $\Sigma_{n\mathbf{k}}'$ is the real part of the electron self-energy defined in Eq.~\eqref{elselfenergy}.
Eq.~\eqref{realpartselfenergy} is motivated by the fact that the Debye-Waller term $\Sigma_{n\mathbf{k}}^{\rm DW}$ is real, static and because the
volume enclosed by the Fermi surface must be conserved after Luttinger's theorem~\cite{Luttinger1960}:
\begin{equation}
 \Re \Sigma_{n\mathbf{k}}(\omega=\varepsilon_{\rm F},T) + \Sigma_{n\mathbf{k}}^{\rm DW}(T) = 0.
\end{equation}

The code can also compute the Fermi surface nesting function defined as~\cite{Bazhirov2010}:
\begin{equation}\label{nestingfct}
\zeta_{\mathbf{q}} = \sum_{nm} \int_{\rm BZ} \frac{d\mathbf{k}}{\Omega_{\rm BZ}} \delta(\varepsilon_{n\mathbf{k}}-\varepsilon_{\rm F}) \delta(\varepsilon_{m\mathbf{k+q}}-\varepsilon_{\rm F}).
\end{equation}

The nesting function is a property of the Fermi surface and is non-zero for wave-vectors that connect two points on the Fermi surface. 

The electron and phonon linewidths can be obtained from the imaginary part of the electron $(\Sigma'')$ and phonon $(\Pi'')$ self-energy~\cite{Giustino2007}:
\begin{multline}\label{linewidthel}
   \Sigma_{n\mathbf{k}}''(\omega,T) = \pi \sum_{m\nu} \int_{\rm BZ} \frac{d\mathbf{q}}{\Omega_{\rm BZ}} | g_{mn,\nu}(\mathbf{k,q}) |^2 \\
    \times \Big\{\big[n_{\mathbf{q}\nu}(T)+f_{m\mathbf{k+q}}(T)\big]\delta(\omega - (\varepsilon_{m\mathbf{k+q}}-\varepsilon_F) +\omega_{\mathbf{q}\nu})\\
+ \big[n_{\mathbf{q}\nu}(T)+1-f_{m\mathbf{k+q}}(T)\big]\delta(\omega - (\varepsilon_{m\mathbf{k+q}}-\varepsilon_F) - \omega_{\mathbf{q}\nu})\Big\}, 
\end{multline}                                        
\begin{multline}\label{linewidthph}
\Pi_{\mathbf{q}\nu}''(\omega,T) = 2\pi \sum_{mn} \int_{\rm BZ} \frac{d\mathbf{k}}{\Omega_{\rm BZ}} | g_{mn,\nu}(\mathbf{k,q}) |^2 \\
\times \big[f_{n\mathbf{k}}(T)-f_{m\mathbf{k+q}}(T)\big]\delta(\varepsilon_{n\mathbf{k}} - \varepsilon_{m\mathbf{k+q}} -\omega).
\end{multline}         
The associated electronic spectral function is:  
\begin{multline}\label{spectral_funct}
   A_{n\mathbf{k}}(\omega,T) =\\
    \frac{1}{\pi} \frac{| \Sigma_{n\mathbf{k}}''(\omega,T)|}{|\omega - (\varepsilon_{n\mathbf{k}}-\varepsilon_F)- \tilde{\Sigma}_{n\mathbf{k}}'(\omega,T)|^2 + | \Sigma_{n\mathbf{k}}''(\omega,T)|}.
\end{multline}
       
Finally, the electron-phonon coupling strength associated with a specific phonon mode $\nu$ and wavevector $\mathbf{q}$ is:
\begin{multline}\label{lambdaEQ}
\lambda_{\mathbf{q}\nu} = \frac{1}{N(\varepsilon_{\rm F}) \omega_{\mathbf{q}\nu}} 
\sum_{nm} \int_{\rm BZ} \frac{d\mathbf{k}}{\Omega_{\rm BZ}} \\
 \times |g_{mn,\nu}(\mathbf{k,q})|^2  \delta(\varepsilon_{n\mathbf{k}}-\varepsilon_{\rm F}) \delta(\varepsilon_{m\mathbf{k+q}}-\varepsilon_{\rm F}),
\end{multline}
where $N(\varepsilon_{\rm F})$ is the density of states per spin at the Fermi level.
From this, the isotropic Eliashberg spectral function can be obtained via an average over the BZ:
\begin{equation}\label{isotropiceliash}
\alpha^2 F(\omega) = \frac{1}{2} \sum_\nu \int_{\rm BZ} \frac{d\mathbf{q}}{\Omega_{\rm BZ}} \omega_{\mathbf{q}\nu} \lambda_{\mathbf{q}\nu} \delta(\omega-\omega_{\mathbf{q}\nu}).
\end{equation}

\section{Electron-phonon coupling in polar materials}\label{polarmat}
In polar materials, electrons can interact strongly with longitudinal optical modes, which give rise to electron-phonon matrix elements  
 $g_{mn,\nu}(\mathbf{k,q})$ that diverge as $1/|\mathbf{q}|$ for $|\mathbf{q}|\rightarrow \mathbf{0}$ \cite{Frohlich1954,Ponce2015}. 
The treatment of such divergence when performing Wannier interpolation has been very recently proposed~\cite{Verdi2015,Sjakste2015}.
The way to tackle this problem is to split the electron-phonon matrix elements into a short- ($\mathcal{S}$) and a long-range ($\mathcal{L}$) contribution~\cite{Verdi2015}: 
\begin{equation}\label{polar_G}
g_{mn,\nu}(\mathbf{k,q}) = g_{mn,\nu}^{\mathcal{S}}(\mathbf{k,q}) + g_{mn,\nu}^{\mathcal{L}}(\mathbf{k,q}),
\end{equation}
where:
\begin{multline}\label{long-range}
g_{mn,\nu}^{\mathcal{L}}(\mathbf{k,q}) = i \sum_\kappa  \bigg( \frac{\hbar}{2 M_\kappa \omega_{\mathbf{q}\nu} } \bigg)^{1/2} \\
\times \sum_{\mathbf{G}\neq -\mathbf{q}} \frac{(\mathbf{q+G})\cdot \mathbf{Z^*}_\kappa \cdot \mathbf{e}_{\kappa,\nu}(\mathbf{q})}{(\mathbf{q+G}) \cdot \boldsymbol{\varepsilon}^{\infty} \cdot (\mathbf{q+G})} \big\langle  \psi_{m\mathbf{k+q}} \big| e^{i(\mathbf{q+G})\cdot \mathbf{r}} \big| \psi_{n\mathbf{k}} \big\rangle.
\end{multline} 
In Eq.~\eqref{long-range}, $\mathbf{Z^*}$ is the Born effective charge tensor, $\mathbf{e}$ is the vibrational eigendisplacement vector and $\boldsymbol{\varepsilon}^{\infty}$ is the macroscopic dielectric permittivity tensor of the electronic system where the ions are considered fixed.  

Most linear-response first-principles codes can compute the full $g_{mn,\nu}(\mathbf{k,q})$ of Eq.~\eqref{polar_G} in polar materials at arbitrary $\mathbf{q}$-points. However, due to the intrinsically localized nature of MLWFs, only the short range component can be treated in a Wannier interpolation scheme. 

The strategy implemented in \texttt{EPW} consists of four steps: (i) the full $g_{mn,\nu}(\mathbf{k,q})$ is computed on a coarse grid using DFPT, (ii) the long-range part $g_{mn,\nu}^{\mathcal{L}}(\mathbf{k,q})$ is then subtracted on that grid to obtain the short-range component $g_{mn,\nu}^{\mathcal{S}}(\mathbf{k,q})$, (iii) the standard Wannier electron-phonon interpolation of Ref.~\cite{Giustino2007} is applied to the short-range component only, and finally (iv) the long-range component is added back to the interpolated short-range part for each arbitrary $\mathbf{k}$ and $\mathbf{q}$-point. 
The overlap matrices in Eq.~\eqref{long-range} can be computed in the approximation of small $\mathbf{q+G}$~\cite{Verdi2015} as:
\begin{equation}
\big\langle \psi_{n\mathbf{k}} \big| e^{i(\mathbf{q+G})\cdot \mathbf{r}} \big| \psi_{m\mathbf{k+q}} \big\rangle = \big[ U_{\mathbf{k}} U_{\mathbf{k+q}}^\dagger \big]_{nm}.
\end{equation}   
The Wannier rotation matrices $U_{nm\mathbf{k}}$ can be obtained at arbitrary $\mathbf{k}$ and $\mathbf{q}$-points through the interpolation of the electronic Hamiltonian~\cite{Souza2001}.

\section{Phonon-mediated superconductors}\label{supercond}

Ab-initio calculations of phonon-mediated superconducting properties are based on the Bardeen-Cooper-Schrieffer (BCS) theory~\cite{Bardeen1957}. There exist three main approaches to the problem: (i) semi-empirical methods based on the McMillan formula~\cite{McMillan1968}; (ii) first-principles Green's function methods based on the Migdal-Eliashberg (ME) theory~\cite{Migdal1958,Eliashberg1960,Eliashberg1961} and; (iii) the density-functional theory for superconductors (SCDFT)~\cite{Luders2005,Marques2005,Linscheid2015,Linscheid2015a}.

Approaches (i) and (ii) are implemented in the \texttt{EPW} code.

\subsection{Allen-Dynes theory} 
The critical temperature $T_{\rm c}$ at which the phase transition occurs can be estimated with semi-empirical methods like the McMillan formula, later refined by Allen and Dynes~\cite{Allen1975} to account for strong electron-phonon coupling:
\begin{equation}\label{ADTc}
T_{\rm c} = \frac{\omega_{\log}}{1.2} \exp\Big(-\frac{1.04(1+\lambda)}{\lambda-\mu_{\rm c}^*(1+0.62\lambda)}\Big),
\end{equation}
where $\omega_{\log}$ is a logarithmic average of the phonon frequency, $\mu_{\rm c}^*$ is the Coulomb pseudopotential, and $\lambda$ is the electron-phonon coupling constant obtained by momentum and mode integration of the electron-phonon coupling strength of Eq.~\eqref{lambdaEQ}.

\subsection{Anisotropic Migdal-Eliashberg theory}\label{eliash}   

\subsubsection{Nambu-Gor'kov generalized Green's function}

The superconducting state of a periodic system can quantitatively be described by the Nambu-Gor'kov generalized Green's function~\cite{Gorkov1958,Nambu1960,Garland1967,Allen1983a,Carbotte1990,Choi2003,Margine2013}:
\begin{equation}\label{eqani1}
\hat{G}_{n\mathbf{k}} (i\omega_j) = 
\begin{bmatrix}
G_{n\mathbf{k}} (i\omega_j) & F_{n\mathbf{k}} (i\omega_j) \\
F_{n\mathbf{k}}^* (i\omega_j) & -G_{n-\mathbf{k}} (-i\omega_j) 
\end{bmatrix},
\end{equation}
where $i\omega_j = i(2j+1)\pi T$ ($j$ integer) are the fermion Matsubara frequencies. 
The Green's function $G$ on the diagonal of Eq.~\eqref{eqani1} describes
 single particle excitations and $F$ is the Gor'kov's anomalous Green's function, which describes the Cooper-pair amplitude. This function is nonzero below $T_{\rm c}$.

Eq.~\eqref{eqani1} can be evaluated by solving Dyson's equation:
\begin{equation}\label{dyson}
\hat{G}_{n\mathbf{k}}^{-1} (i\omega_j) = \hat{G}_{n\mathbf{k}}^{\text{no}^{-1}} (i\omega_j)  - \hat{\Sigma}_{n\mathbf{k}}^{\text{pa}} (i\omega_j), 
\end{equation}
where $\hat{G}^{\text{no}}$ is the non-interacting electron Green's function, and $\hat{\Sigma}^{\text{pa}}$ is the self-energy describing the pairing interaction.  

The two components of Eq.~\eqref{dyson} can be expressed using a linear combination of Pauli matrices:
\begin{equation}\label{noninteractingG}
\hat{G}_{n\mathbf{k}}^{\text{no}^{-1}} (i\omega_j)  = i\omega_j \hat{\tau}_0 - (\varepsilon_{n\bf{k}}-\varepsilon_{\rm F})\hat{\tau}_3,
\end{equation}
and:
\begin{multline}\label{selfsigma}
\hat{\Sigma}_{n\mathbf{k}}^{\text{pa}} (i\omega_j) = i\omega_j [1-Z_{n\mathbf{k}}(i\omega_j)]\hat{\tau}_0 + \chi_{n\bf{k}}(i\omega_j)\hat{\tau}_3\\
+\phi_{n\bf{k}}(i\omega_j)\hat{\tau}_1  +\bar{\phi}_{n\bf{k}}(i\omega_j)\hat{\tau}_2,
\end{multline}
where the mass renormalization function $Z$, the energy shift $\chi$, and the order parameter $\phi$ are scalar functions to be determined. 
The Pauli matrices are given by:
\begin{equation}
\hat{\tau}_0 = \begin{bmatrix}1 & 0 \\
0 & 1 
\end{bmatrix},\hat{\tau}_1 = \begin{bmatrix}0 & 1 \\
1 & 0 
\end{bmatrix},\hat{\tau}_2 = \begin{bmatrix}0 & -i \\
i & 0 
\end{bmatrix},\hat{\tau}_3 = \begin{bmatrix}1 & 0 \\
0 & -1
\end{bmatrix}.
\end{equation}

Inserting Eqs.~\eqref{noninteractingG} and \eqref{selfsigma} into Eq.~\eqref{dyson} and inverting the matrix leads to: 
\begin{multline}\label{gnormaleq}
\hat{G}_{n\mathbf{k}}(i\omega_j)  =\frac{-1}{\theta_{n\mathbf{k}}(i\omega_j)}\Big\{ i\omega_j Z_{n\bf{k}}(i\omega_j)\hat{\tau}_0 \\
+\big[\varepsilon_{n\bf{k}}-\varepsilon_{\rm F} +\chi_{n\bf{k}}(i\omega_j)\big]\hat{\tau}_3  
+ \phi_{n\bf{k}}(i\omega_j)\hat{\tau}_1 + \bar{\phi}_{n\bf{k}}(i\omega_j)\hat{\tau}_2  \Big\},
\end{multline}
where $-\theta_{n\mathbf{k}}(i\omega_j)$ is the determinant of $\hat{G}_{n\mathbf{k}}^{-1} (i\omega_j)$:
\begin{multline}\label{defofdeno}
\theta_{n\mathbf{k}}(i\omega_j)  = \big[\omega_j Z_{n\bf{k}}(i\omega_j)\big]^2 \\
+ \big[ \varepsilon_{n\bf{k}}-\varepsilon_{\rm F} + \chi_{n\bf{k}}(i\omega_j)\big]^2 
+ \phi_{n\bf{k}}^2(i\omega_j) + \bar{\phi}_{n\bf{k}}^2(i\omega_j).
\end{multline}

In the normal state, $Z$ and $\chi$ can be expressed as a linear combination of the self-energy $\hat{\Sigma}^{\text{pa}}$~\cite{Allen1983a}:
\begin{align}
i\omega_j\big[1-Z_{n\bf{k}}(i\omega_j) \big] &= 1/2 \big[ \hat{\Sigma}_{n\mathbf{k}}^{\text{pa}} (i\omega_j)  - \hat{\Sigma}_{n\mathbf{k}}^{\text{pa}} (-i\omega_j) \big] \\
\chi_{n\bf{k}}(i\omega_j) &= 1/2 \big[ \hat{\Sigma}_{n\mathbf{k}}^{\text{pa}} (i\omega_j)  + \hat{\Sigma}_{n\mathbf{k}}^{\text{pa}} (-i\omega_j) \big], 
\end{align}
where it has been assumed that $Z$ and $\chi$ are both even functions of $\bf{k}$ and $\omega_j$.
Owing to the relationship between the Green's functions on the diagonal of Eq.~\eqref{eqani1}, $\hat{G}_{n\bf{k}}(i\omega_j)=-\hat{G}_{n-\bf{k}}(-i\omega_j)$,  Eq.~\eqref{gnormaleq} leads to: 
\begin{equation}
\phi_{n\bf{k}}^2(i\omega_j)+\bar{\phi}_{n\bf{k}}^2(i\omega_j)  = \phi_{n-\bf{k}}^2(-i\omega_j) + \bar{\phi}_{n-\bf{k}}^2(-i\omega_j), 
\end{equation}
which implies that $\phi$ and $\bar{\phi}$ are also even functions of $\bf{k}$ and $\omega_j$.
Furthermore, due to the form of the order parameters in Eq.~\eqref{selfsigma}, it is expected that $\phi$ and $\bar{\phi}$ will be proportional: 
\begin{equation}
(\phi,\bar{\phi}) = \big(\phi_0 \cos \alpha, \phi_0 \sin \alpha \big).
\end{equation}
Without loss of generality, one can therefore choose the gauge $\alpha=0$, which leads to $\bar{\phi}=0$. 

\subsubsection{Migdal-Eliashberg approximation}
\begin{figure}[b]
  \centering
  \includegraphics[width=0.99\linewidth]{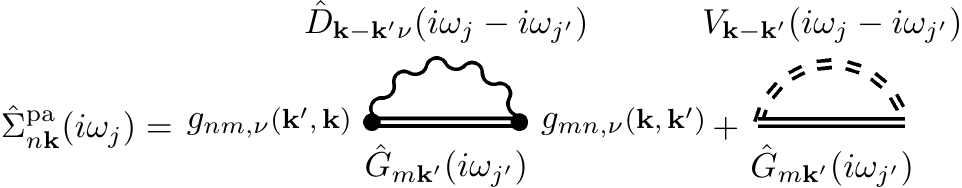}
  \caption{\label{fan} Feynman diagrams of the pairing self-energy $\hat{\Sigma}_{n\mathbf{k}}^{\text{pa}} (i\omega_j)$, within the Migdal approximation.}
\end{figure}

The Eliashberg theory is a generalization of the BCS theory to include retardation effects~\cite{Eliashberg1960,Eliashberg1961}.
In the formulation of Allen and Mitrovi\'c~\cite{Allen1983a} the pairing self-energy is written as the sum of an electron-phonon and a screened Coulomb interaction:
\begin{equation}\label{sumephCoulb}
\hat{\Sigma}_{n\mathbf{k}}^{\text{pa}} (i\omega_j) = \hat{\Sigma}_{n\mathbf{k}}^{\text{ep}} (i\omega_j)+ \hat{\Sigma}_{n\mathbf{k}}^{\text{c}} (i\omega_j),
\end{equation}
with:
\begin{multline}\label{elphselfenerg}
\hat{\Sigma}_{n\mathbf{k}}^{\text{ep}}(i\omega_j) = -T \sum_{m\mathbf{k'}j'\nu} \hat{\tau}_3 \hat{G}_{m\bf{k'}}(i\omega_{j'})\hat{\tau}_3 \\
\times |g_{mn,\nu}(\mathbf{k,k'})|^2 D_{\mathbf{k-k'}\nu}(i\omega_j-i\omega_{j'}),
\end{multline}
and: 
\begin{multline}\label{coulombself}
\hat{\Sigma}_{n\mathbf{k}}^{\text{c}} (i\omega_j) =\\
 -T \sum_{m\bf{k'}j'}  \hat{\tau}_3 \hat{G}_{m\bf{k'}}(i\omega_{j'})\hat{\tau}_3  V_{n\mathbf{k}-m\mathbf{k'}}(i\omega_j-i\omega_{j'}).
\end{multline}
Here $V$ is the dynamically screened Coulomb interaction between electrons, and $D$ is the dressed phonon propagator. The two Feynman diagrams associated with Eq.~\eqref{sumephCoulb} are given in Fig.~\ref{fan}.

Note that the two self-energies of Eqs.~\eqref{elphselfenerg} and \eqref{coulombself} only contain the leading Feynman diagram. The approximation that allows one to neglect all other Feynman diagrams is called the Migdal theorem~\cite{Migdal1958} and rests on the observation that the neglected terms are of the order of the square-root of the electron to ion mass ratio. 
We also rely on the band-diagonal approximation~\cite{Allen1978a,Chakraborty1978a,Pickett1982}, that neglects band-mixing (i.e. an additional band index) in $G$ and $\Sigma$. Since the superconducting energy pairing is very small, such approximation should be very good for non-degenerate bands.

The dressed phonon propagator can be expressed in terms of its spectral representation as~\cite{Allen1983a,Marsiglio2008}: 
\begin{equation}\label{phononpropa}
D_{\bf{q}\nu}(i\omega_l) =-\int_0^{\infty} d \Omega B_{\bf{q}\nu}(\Omega)\frac{2\Omega}{\omega_l^2+\Omega^2},
\end{equation}
where $i\omega_l=i 2\pi  l T$ ($l$ integer) is the boson Matsubara frequency and $B$ is the phonon spectral function, that is the phonon density of states. 
Inserting Eq.~\eqref{phononpropa} into Eq.~\eqref{elphselfenerg} leads to: 
\begin{multline}\label{elphselfenerg2}
\hat{\Sigma}_{n\mathbf{k}}^{\text{ep}}(i\omega_j) = T \sum_{m\mathbf{k'}j'\nu}\int_0^{\infty} d \Omega \hat{\tau}_3 \hat{G}_{m\bf{k'}}(i\omega_{j'})\hat{\tau}_3 \\
\times |g_{mn,\nu}(\mathbf{k,k'})|^2  B_{\bf{k-k'}\nu}(\Omega)\frac{2\Omega}{(\omega_j-\omega_{j'})^2+\Omega^2}.
\end{multline}

The form of Eq.~\eqref{elphselfenerg2} allows one to introduce the Eliashberg electron-phonon spectral function as follows:
\begin{multline}\label{Elishberga2F}
\alpha^2 F(n\mathbf{k},m \mathbf{k'},\Omega) \\
= N(\varepsilon_{\rm F}) \sum_\nu |g_{mn,\nu}(\mathbf{k,k'})|^2 B_{\bf{k-k'}\nu}(\Omega),
\end{multline}
where $N(\varepsilon_{\rm F})$ is the electron density of states at the Fermi level.
To compare with available experimental data, it is worthwhile to define Fermi-surface-averaged spectral functions:
\begin{multline}
\alpha^2 F(n\mathbf{k},\Omega) \\
= \frac{1}{N(\varepsilon_{\rm F})} \sum_{m\bf{k'}} \alpha^2 F(n\mathbf{k},m \mathbf{k'},\Omega) \delta(\varepsilon_{m\bf{k'}}-\varepsilon_{\rm F}),
\end{multline}
and:
\begin{equation}
\alpha^2 F(\Omega) = \frac{1}{N(\varepsilon_{\rm F})} \sum_{n\bf{k}} \alpha^2 F(n\mathbf{k},\Omega) \delta(\varepsilon_{n\bf{k}}-\varepsilon_{\rm F}).
\end{equation}

The rationale for performing the averages around the Fermi surface is that the phonon energy is usually much smaller than the Fermi energy. However,
the fully anisotropic $\alpha^2 F(n\mathbf{k},m \mathbf{k'},\Omega)$ has been implemented into the \texttt{EPW} software in order to fully take into account multi-band effects and systems of reduced dimensionality.

Now, inserting Eqs.~\eqref{coulombself}, \eqref{elphselfenerg2} and \eqref{Elishberga2F} into Eq.~\eqref{sumephCoulb}, using Eq.~\eqref{gnormaleq} with $\bar{\phi}=0$ to evaluate $\hat{G}_{n\mathbf{k}}(i\omega_j)$, 
and multiplying the Pauli matrices together lead to: 
\begin{multline}\label{eqlong}
\hat{\Sigma}_{n\mathbf{k}}^{\text{pa}} (i\omega_j) = -T \sum_{m\mathbf{k'}j'} \frac{1}{\theta_{m\mathbf{k'}}(i\omega_{j'})}\Big\{ i\omega_j Z_{m\bf{k'}}(i\omega_{j'})\hat{\tau}_0 \\
+[\varepsilon_{m\bf{k'}}-\varepsilon_{\rm F} +\chi_{m\bf{k'}}(i\omega_{j'})]\hat{\tau}_3  - \phi_{m\bf{k'}}(i\omega_{j'})\hat{\tau}_1 \Big\} \\
\times \Bigg\{\frac{\lambda(n\mathbf{k},m\mathbf{k'},\omega_j-\omega_{j'})}{N(\varepsilon_{\rm F})} -  V_{n\mathbf{k}-m\mathbf{k'}}(i\omega_j-i\omega_{j'})\Bigg\},
\end{multline}
where $\theta$ is the same as Eq.~\eqref{defofdeno} but with $\bar{\phi}=0$, and where we have introduced the anisotropic electron-phonon coupling strength:
\begin{multline}\label{couplingstrengthbase}
    \lambda(n\mathbf{k},m\mathbf{k'},\omega_j-\omega_{j'}) =\\
     \int_0^{\infty} d\Omega \frac{2\Omega}{(\omega_j-\omega_{j'})^2+\Omega^2} \alpha^2 F(n\mathbf{k},m\mathbf{k'},\Omega). 
\end{multline}  

Direct comparison of the different Pauli matrix components of Eq.~\eqref{eqlong} with Eq.~\eqref{selfsigma} leads to a system of three coupled non-linear equations:
\begin{multline}\label{firstequ}
Z_{n\mathbf{k}}(i\omega_j) = 1 + \frac{T}{\omega_j} \sum_{m\mathbf{k'}j'} \frac{\omega_{j'}Z_{m\mathbf{k'}}(i\omega_{j'})}{\theta_{m\mathbf{k'}}(i\omega_{j'})}\\
\times \Bigg\{\frac{\lambda(n\mathbf{k},m\mathbf{k'},\omega_j-\omega_{j'})}{N(\varepsilon_{\rm F})} -  V_{n\mathbf{k}-m\mathbf{k'}}(i\omega_j-i\omega_{j'})\Bigg\},
\end{multline}
\begin{multline}\label{secondeq}
\chi_{n\mathbf{k}}(i\omega_j) = -T \sum_{m\mathbf{k'}j'} \frac{\varepsilon_{m\mathbf{k'}}-\varepsilon_{\rm F}+\chi_{m\mathbf{k'}}(i\omega_{j'})}{\theta_{m\mathbf{k'}}(i\omega_{j'})}\\
\times \Bigg\{\frac{\lambda(n\mathbf{k},m\mathbf{k'},\omega_j-\omega_{j'})}{N(\varepsilon_{\rm F})} -  V_{n\mathbf{k}-m\mathbf{k'}}(i\omega_j-i\omega_{j'})\Bigg\},
\end{multline}
and
\begin{multline}\label{thirdone}
\phi_{n\mathbf{k}}(i\omega_j) = T \sum_{m\mathbf{k'}j'} \frac{\phi_{m\mathbf{k'}}(i\omega_{j'})}{\theta_{m\mathbf{k'}}(i\omega_{j'})}\\
\times \Bigg\{\frac{\lambda(n\mathbf{k},m\mathbf{k'},\omega_j-\omega_{j'})}{N(\varepsilon_{\rm F})} -  V_{n\mathbf{k}-m\mathbf{k'}}(i\omega_j-i\omega_{j'})\Bigg\}.
\end{multline}

This set of equations needs to be supplemented with an equation for the electron number $N_{\rm e}$ in order to determine the Fermi energy~\cite{Marsiglio2008}:
\begin{equation}
N_{\rm e} = 1 - 2T \sum_{m\mathbf{k'}j'}  \frac{\varepsilon_{m\mathbf{k'}}-\varepsilon_{\rm F}+\chi_{m\mathbf{k'}}(i\omega_{j'})}{\theta_{m\mathbf{k'}}(i\omega_{j'})}.
\end{equation}

The superconducting gap can then be obtained as the ratio between the order parameter and the renormalization function:
\begin{equation}
\Delta_{n\mathbf{k}}(i\omega_{j}) = \frac{\phi_{n\mathbf{k}}(i\omega_j)}{Z_{n\mathbf{k}}(i\omega_j)}.
\end{equation}
The highest temperature for which these coupled equations admit non trivial solutions $\Delta_{n\mathbf{k}}(i\omega_{j}) \neq 0$ defines the critical temperature $T_{\rm c}$.

\subsubsection{Standard approximations}

In order to solve these equations, it is common practice to make the following approximations:  
(i) only the off-diagonal contribution $(\hat{G}^{od})$ to the Green's function is considered in Eq.~\eqref{coulombself} in order to avoid double counting of Coulomb effects~\cite{Allen1983a};
(ii) static screening approximation~\cite{Scalapino1966};
(iii) all quantities are evaluated around the Fermi surface~\cite{Scalapino1966,Scalapino1969,Allen1976a,Allen1983a,Choi2003,Marsiglio2008,Margine2013}. This approximation stems from the fact that the superconducting pairing occurs mainly in an energy window of the size of a phonon energy around the Fermi level; 
(iv) the electron density of states is assumed to be constant and the bandwidth is infinite~\cite{Marsiglio2008}; and
(v) the dynamically screened Coulomb interaction $N(\varepsilon_{\rm F}) V_{n\mathbf{k}-m\mathbf{k'}}$ is embedded into the semiempirical Morel-Anderson pseudopotential $\mu_{\rm c}^*$~\cite{Morel1962}. For a wide range of superconductors, a value of $\mu_{\rm c}^*$ in the range 0.1-0.2 has been found. 
Very recently \texttt{EPW} has been used successfully together with a $\mu_{\rm c}^*$ calculated from first principles~\cite{Margine2016}. 

As a consequence of approximation (ii), the Coulomb contribution to $Z_{n\mathbf{k}}(i\omega_j)$ vanishes because the $\hat{\tau}_0$ component of $\hat{G}_{n\bf{k}}(i\omega_j)$ is an odd function of $\omega_j$~\cite{Scalapino1966}.

Following approximation (iii), the identity $\int_{-\infty}^{\infty} d\varepsilon \delta(\varepsilon_{m\mathbf{k}}-\varepsilon_{\rm F} -\varepsilon)=1$ can be inserted into Eqs.~\eqref{firstequ}-\eqref{thirdone}. 
The resulting equations can be integrated using contour integration in the upper-half plane with the poles of the denominator $\theta$ being given by:
\begin{equation}
\varepsilon+\chi_{m\mathbf{k'}}(i\omega_{j'})= \pm i \sqrt{\omega_{j'}^2 Z_{m\mathbf{k'}}^2(\omega_{j'}) + \phi_{m\mathbf{k'}}^2(\omega_{j'})}.
\end{equation}

Approximation (iv) implies that $\chi = 0$ as Eq.~\eqref{secondeq} is odd in energy and therefore vanishes when integrated.
These four approximations lead to the two nonlinear coupled equations (to be solved self-consistently)~\cite{Choi2003,Margine2013}:  
\begin{multline}\label{coupledAnisoEqs1}
   Z_{n\mathbf{k}}(i\omega_j) =  1 + \frac{\pi T}{N(\varepsilon_{\rm F})\omega_j} \sum_{m\mathbf{k'}j'}  \frac{\omega_{j'}}{\sqrt{\omega_{j'}^2+\Delta_{m\mathbf{k'}}^2(i\omega_{j'})} }\\
   \times \lambda(n\mathbf{k},m\mathbf{k'},\omega_{j}-\omega_{j'})\delta(\epsilon_{m\mathbf{k'}}-\varepsilon_{\rm F}), 
\end{multline}                                      
\begin{multline}\label{coupledAnisoEqs2}
    Z_{n\mathbf{k}}(i\omega_j)\Delta_{n\mathbf{k}}(i\omega_{j}) = \frac{\pi T}{N(\varepsilon_{\rm F})}  \sum_{m\mathbf{k'}j'} \frac{\Delta_{m\mathbf{k'}}(i\omega_{j'})}{\sqrt{\omega_{j'}^2+\Delta_{m\mathbf{k'}}^2(i\omega_{j'})} } \\
   \times [\lambda(n\mathbf{k},m\mathbf{k'},\omega_j-\omega_{j'})-\mu_{\rm c}^*]\delta(\epsilon_{m\mathbf{k'}}-\varepsilon_{\rm F}).
\end{multline}          

An important observable related to superconductivity that can directly be computed is the superconducting specific heat. The specific heat can be obtained from
the free energy difference between the superconducting and normal states~\cite{Bardeen1964,Choi2003,Marsiglio2008}:
\begin{multline}\label{free_energy}
\Delta F = -\pi T \sum_{ n\mathbf{k}j} \bigg[ \sqrt{\omega_j^2 + \Delta_{n\mathbf{k}}^2(i\omega_j)} - |\omega_j| \bigg]\\
\times \Bigg[ Z_{n\mathbf{k}}(i\omega_j) - Z_{n\mathbf{k}}^N(i\omega_j)\frac{|\omega_j|}{\sqrt{\omega_j^2 + \Delta_{n\mathbf{k}}^2(i\omega_j)} }  \Bigg]\delta(\varepsilon_{n\mathbf{k}}-\varepsilon_{\rm F}),
\end{multline}
where $Z^N$ is the normal-state $Z$, which is obtained from Eq.~\eqref{coupledAnisoEqs1} by setting $\Delta = 0$.
From this, the specific heat difference can be obtained as:
\begin{equation}\label{specific_heat}
\Delta C(T) = -T \frac{d^2 \Delta F}{dT^2}.
\end{equation}
The specific heat in the superconducting state raises exponentially with temperature at low temperature, and then undergoes a characteristic discontinuous jump at the critical temperature $T_{\rm c}$~\cite{Phillips1959}. 

Eqs.~\eqref{coupledAnisoEqs1} and \eqref{coupledAnisoEqs2} can be efficiently solved on the imaginary axis. However, physically relevant properties like tunneling current and heat capacity require the knowledge of the superconductivity gap along the real axis~\cite{McMillan1965}.

As the direct evaluation of the Eliashberg equations on the real axis is computationally demanding~\cite{Schrieffer1963,Scalapino1966,Holcomb1996} we choose to obtain the solution by analytic continuation from the imaginary axis to the real axis. The \texttt{EPW} software supports analytic continuation using Pad\'e approximants~\cite{Vidberg1977,Leavens1985} and the exact iterative procedure of Ref.~\cite{Marsiglio1988}. More information can be found in Ref.~\cite{Margine2013}.   

From the knowledge of the mass renormalization $Z_{n\mathbf{k}}(\omega)$ and superconducting gap $\Delta_{n\mathbf{k}}(\omega)$ on the real axis, we obtain the diagonal components of the single electron Green's function~\cite{Scalapino1969,Marsiglio2008}:
\begin{equation}\label{diagonalGreen}
G_{n\mathbf{k}}(\omega) = \frac{\omega Z_{n\mathbf{k}}(\omega)+(\varepsilon_{n\mathbf{k}}-\varepsilon_F)}{\omega^2 Z_{n\mathbf{k}}^2(\omega)-(\varepsilon_{n\mathbf{k}}-\varepsilon_F)^2-Z_{n\mathbf{k}}^2(\omega)\Delta_{n\mathbf{k}}^2(\omega)}.
\end{equation}

Eq.~\eqref{diagonalGreen} has poles located at:
\begin{equation}
\tilde{E}_{n\mathbf{k}}^2 =  \Big[E_{n\mathbf{k}} -i(\tau_{n\mathbf{k}}^{\rm su})^{-1}\Big]^2 = \frac{(\varepsilon_{n\mathbf{k}}-\varepsilon_F)^2}{Z_{n\mathbf{k}}^2(\tilde{E}_{n\mathbf{k}})} +  \Delta_{n\mathbf{k}}^2(\tilde{E}_{n\mathbf{k}}), 
\end{equation}
where the real part of $\tilde{E}_{n\mathbf{k}}$, $E_{n\mathbf{k}}$, is the quasiparticle energy renormalized by the superconducting pairing, and   the imaginary part of  $\tilde{E}_{n\mathbf{k}}$, $(\tau_{n\mathbf{k}}^{\rm su})^{-1}$, represents the scattering time due to  the superconducting pairing.
Such renormalized quasiparticle energies  $E_{n\mathbf{k}}$ have been recently studied (e.g. in intercalated graphite~\cite{Sanna2012}).

Introducing the real and imaginary part of $Z_{n\mathbf{k}}=Z_{1n\mathbf{k}}+iZ_{2n\mathbf{k}}$ and $\Delta_{n\mathbf{k}}=\Delta_{1n\mathbf{k}}+i\Delta_{2n\mathbf{k}}$ and linearizing the imaginary part (i.e. we neglect $\Delta_{2n\mathbf{k}}^2$, $Z_{2n\mathbf{k}}^2$, $\Delta_{2n\mathbf{k}}Z_{2n\mathbf{k}}$ and $(\tau_{n\mathbf{k}}^{\rm su})^2$ terms) lead to the following expression for the quasiparticle energy~\cite{Kaplan1976,Marsiglio1997,Marsiglio1997a}: 
\begin{equation}\label{gapequation}
E_{n\mathbf{k}} = \sqrt{\frac{(\varepsilon_{n\mathbf{k}}-\varepsilon_F)^2}{Z_{1n\mathbf{k}}^2(E_{n\mathbf{k}})}+\Delta_{1n\mathbf{k}}^2(E_{n\mathbf{k}})},
\end{equation}
and its associated lifetime:
\begin{equation}
\frac{1}{\tau_{n\mathbf{k}}^{\rm su}} = \frac{Z_{2n\mathbf{k}}}{Z_{1n\mathbf{k}}}\times \frac{\frac{(\varepsilon_{n\mathbf{k}}-\varepsilon_F)^2}{Z_{1n\mathbf{k}}^2}+\Delta_{1n\mathbf{k}}^2-\Delta_{1n\mathbf{k}}\Delta_{2n\mathbf{k}}}{\sqrt{\frac{(\varepsilon_{n\mathbf{k}}-\varepsilon_F)^2}{Z_{1n\mathbf{k}}^2}+\Delta_{1n\mathbf{k}}^2}}.
\end{equation}
Although lifetimes are not currently directly implemented in \texttt{EPW}, they can be computed as post-processing.

Finally we can deduce the theoretical tunneling density of states~(\cite{Schrieffer1964}, p.190):
\begin{equation}\label{tunnelingdensity}
\frac{N_S(\omega)}{N(\varepsilon_{\rm F})} = \frac{-1}{\pi}\int_{-\infty}^{\infty} d\varepsilon_{n\mathbf{k}} \Im G_{n\mathbf{k}}(\omega),
\end{equation}
where $N_S(\omega)$ is the superconducting quasiparticle density of states for the quasiparticle energy $\omega$ and $N(\varepsilon_{\rm F})$ is the normal density of states at the Fermi level. 
                                                                                                                                                                                                                                                                                                                                                                                                                                                                                                                                                                                                                                                                                                                                       
                                                                                                                                                                                                                                                                                                                                                                                                                                                                                                                                                                                                                                                                                                                                       In the BCS limit, $Z=1$ and $\Delta$ is considered to be independent of band index and wavevector. Within this limit, Eq.~\eqref{tunnelingdensity} can be approximated using Eq.~\eqref{diagonalGreen}; integrating the two poles using the residues theorem leads to:
\begin{equation}\label{tunnelingdensity2}
\frac{N_S(\omega)}{N(\varepsilon_{\rm F})} = \Re  \frac{\omega}{\sqrt{\omega^2 - \Delta^2(\omega)}}.
\end{equation}

\section{Transport properties}\label{transportProp}

Electronic transport properties in metals and doped semiconductors can be calculated by solving the Boltzmann transport equation (BTE)~\cite{Ziman1960}.
Several approximations for the solution of the BTE have been proposed. 
Here we consider the simplest and most popular approximation: the Ziman's resistivity formula. This formula is based on an average of the Eliashberg coupling function (see Eq. 9.5.22 of Ref.~\cite{Ziman1960} or Eq.~8.21 of Ref.~\cite{Grimvall1981}):
\begin{multline}\label{resistivitytr}
\rho(T) = \frac{4\pi m_e}{n e^2 k_B T}\\
\times \int_0^{\infty} \! d\omega\, \hbar \omega\, \alpha_{\text{tr}}^2 F(\omega)\, n(\omega,T)\big[1+n(\omega,T)\big],
\end{multline}
where $n(\omega,T)$ is the Bose-Einstein distribution. We have introduced in Eq.~\eqref{resistivitytr} the Eliashberg transport coupling function:
\begin{equation}\label{eliashtransport}
\alpha_{\text{tr}}^2 F(\omega) =  \frac{1}{2} \sum_{\nu}\int_{\rm BZ} \frac{d\mathbf{q}}{\Omega_{\rm BZ}} \omega_{\mathbf{q}\nu} \lambda_{\text{tr},\mathbf{q}\nu} \delta(\omega-\omega_{\mathbf{q}\nu}),
\end{equation} 
where the mode-resolved coupling strength is analogous to Eq.~\eqref{couplingstrengthbase} and is defined by: 
\begin{multline}\label{Zimantransport}
\lambda_{\text{tr},\mathbf{q}\nu} = \frac{1}{N(\varepsilon_F) \omega_{\mathbf{q}\nu}}  \sum_{nm}  \int_{\rm BZ} \frac{d\mathbf{k}}{\Omega_{\rm BZ}} |g_{mn,\nu}(\mathbf{k,q})|^2 \\
 \times \delta(\varepsilon_{n\mathbf{k}}-\varepsilon_{\rm F}) \delta(\varepsilon_{m\mathbf{k+q}}-\varepsilon_{\rm F})\Big( 1-\frac{v_{n\mathbf{k}}\cdot v_{m\mathbf{k+q}}}{|v_{n\mathbf{k}}|^2}\Big).
\end{multline}
Here $v_{n\mathbf{k}} = \partial \varepsilon_{n\mathbf{k}}/\partial \mathbf{k}$ is the electron velocity. A slightly modified version of Eq.~\eqref{Zimantransport} that conserves the norm of the velocity term could be used with a factor:
\begin{equation}
\Big( 1-\frac{v_{n\mathbf{k}}\cdot v_{m\mathbf{k+q}}}{|v_{n\mathbf{k}}||v_{m\mathbf{k+q}}|}\Big).
\end{equation}
 Note that Eq.~\eqref{Zimantransport} relies on a semi-classical approximation where the velocities are expectation values of single electron velocities. A more general framework should rely on the matrix elements $v_{mn\mathbf{k}}$, see Eq.~\eqref{finalvelo} below.
Extensions to the anisotropic case can easily be implemented in the \texttt{EPW} software, given the availability of the anisotropic Eliashberg function of Eq.~\eqref{Elishberga2F}, but are left for future work. 

It is worth mentioning that the calculation of the the electron velocity can be performed using different techniques: (i) finite differences; (ii) analytic derivatives in the \textit{local approximation}; (iii) analytic derivatives within perturbation theory~\cite{Ashcroft1976,Yates2007,Janssen2016}. 
The \texttt{EPW} software currently supports (i) and (ii), but efforts will be dedicated in the future to integrate the Wannier formalism of (iii) into the software following Ref.~\cite{Yates2007} and using routines from the \texttt{wannier90}~\cite{Mostofi2008} code.

\section{Electronic velocities}\label{veloelectro}

The periodic velocity operator can be expressed in terms of the commutator between the Hamiltonian of the system and the position operator, $\hat{v}_\alpha = i[\hat{H},\hat{r}_\alpha]$, with $\alpha=1,2,3$ indicating the Cartesian direction. The matrix elements of $\hat{v}_\alpha$ are~\cite{Starace1971,Read1991,Ismail-Beigi2001,Pickard2003}:
\begin{equation}\label{basicvelo}
v_{mn\mathbf{kk'}\alpha} = \langle  \psi_{m\mathbf{k'}}  | \hat{v}_\alpha | \psi_{n\mathbf{k}} \rangle = \langle \psi_{m\mathbf{k'}}  | \hat{p}_\alpha +i[\hat{V}_{\rm NL},\hat{r}_\alpha] | \psi_{n\mathbf{k}}  \rangle, 
\end{equation}
where $\hat{p}_\alpha = -i \nabla_\alpha$ is the momentum operator and $\hat{V}_{\rm NL}$ is the nonlocal pseudopotential. 
Neglecting the term with $\hat{V}_{\rm NL}$ in the case of a nonlocal pseudopotential is what we call the \textit{local approximation}. Within this approximation,  Eq.~\eqref{basicvelo} reads:  
\begin{multline}\label{eqvelocity}
v_{mn\mathbf{kk'}\alpha}  \approx \langle \psi_{m\mathbf{k'}} | \hat{p}_\alpha |  \psi_{n\mathbf{k}}  \rangle   = \delta(\mathbf{k-k'})\\
\times \bigg( k_\alpha \delta_{mn} - i\int d\mathbf{r} u_{m\mathbf{k'}}^*(\mathbf{r}) \nabla_\alpha u_{n\mathbf{k}}(\mathbf{r}) \bigg),
\end{multline}

If the second part of Eq.~\eqref{eqvelocity} is expanded into plane-waves $u_{n\mathbf{k}}= \sum_{\mathbf{G}} c_{n\mathbf{k}}(\mathbf{G})e^{i\mathbf{G}\cdot \mathbf{r}}$, we obtain the desired velocity expression in the \textit{local approximation} $\tilde{v}$:
\begin{equation}\label{finalvelo}
\tilde{v}_{mn\mathbf{k}} =   \mathbf{k} \delta_{mn} + \sum_{\mathbf{G}} c_{m\mathbf{k}}(\mathbf{G})^*c_{n\mathbf{k}}(\mathbf{G})\mathbf{G} .
\end{equation}

We note that this expression is only valid in the case of norm-conserving pseudopotentials. 
To overcome the \textit{local approximation}, we would need to evaluate the matrix elements $\langle \psi_{m\mathbf{k'}}  | \hat{V}_{\rm NL}\hat{r}_\alpha- \hat{r}_\alpha \hat{V}_{\rm NL} | \psi_{n\mathbf{k}}  \rangle$. This is certainly doable but is left for future work. Alternatively, the velocities can be expressed as first-order derivatives of the Hamiltonian and treated analytically~\cite{Blount1962,Yates2007,Janssen2016}.

\section{Parallelization and Speedup}\label{para}

Considerable effort was devoted to the optimization of the code. Part of the code was re-written in a more efficient way, focusing on the extended use of freely-available external library software packages like Basic Linear Algebra Subprograms (BLAS) and Linear Algebra PACKage (LAPACK).

\begin{figure}[b!]
  \centering
  \includegraphics[width=0.9\linewidth]{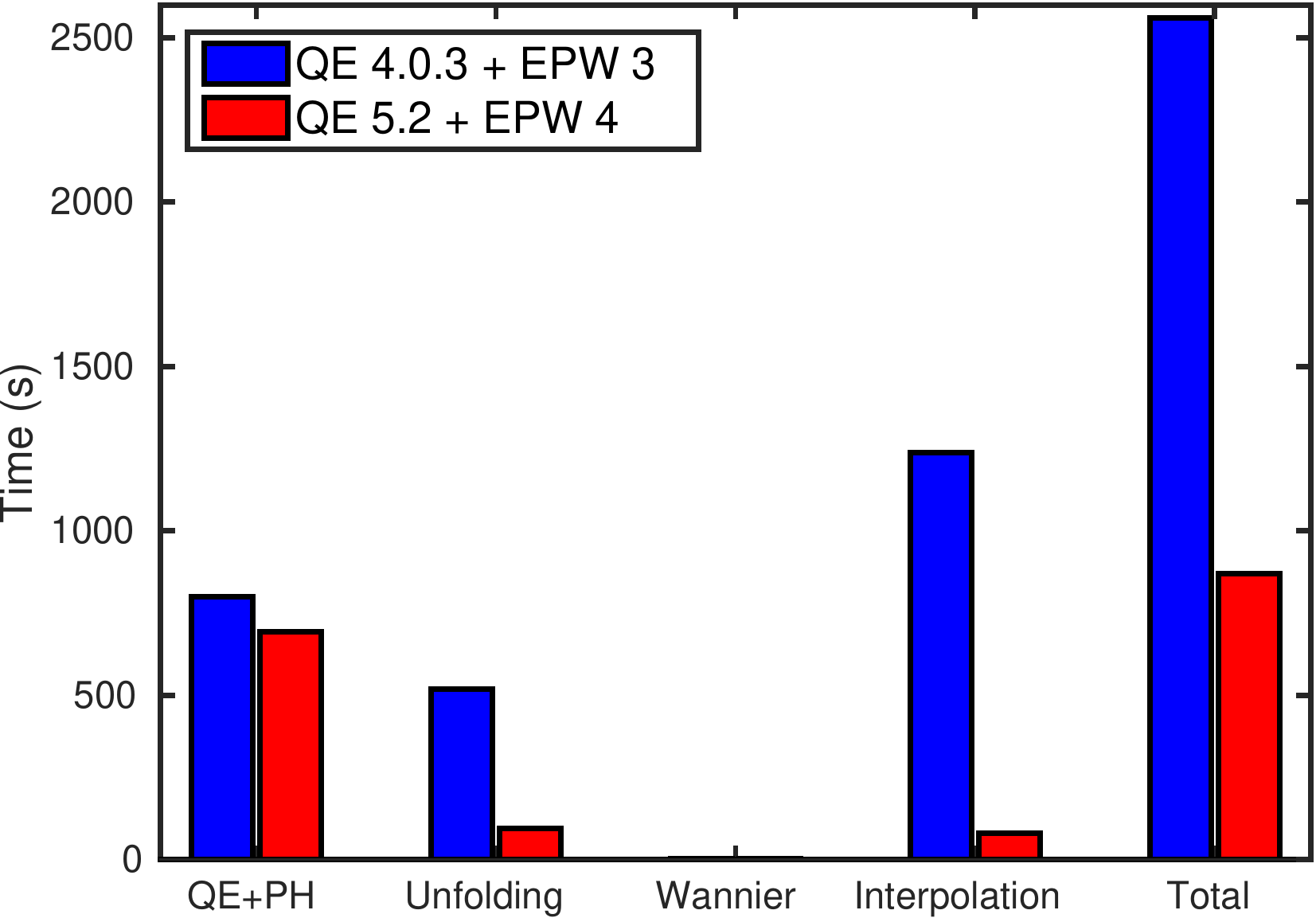}
  \caption{\label{Speedup}(Color online) Comparison of the time required to compute the electronic lifetime of SiC using \texttt{EPW} 3 and the new version 4 of \texttt{EPW}, run on one processor. We show the time required for the calculation of the electrons and phonons perturbations using DFPT (QE+PH), the calculations of the electron-phonon matrix elements and their unfolding from the IBZ to the BZ using the crystal symmetries (Unfolding), the Wannierization from the coarse Bloch space to the real space (Wannier) and the interpolation from real space to fine grids in Bloch space (Interpolation). }
\end{figure}

Such effort lead to a global speedup of 294\% between the current and previous version of the codes, as shown in Fig.~\ref{Speedup}.
The numerical results differ by less than 6\% in relative difference on sensitive physical quantities like phonon frequencies, self-energies, electron-phonon coupling strengths or Eliashberg spectral functions. 

For this test we used SiC as an example, with a lattice parameter of 8.237~bohr, $3\times3\times3$ $\Gamma$-centered $\mathbf{k}$-point grid, and 60~Ry for the plane-wave energy cutoff. The fine grids on which the Wannier interpolation was performed are a $10\times10\times10$ $\mathbf{k}$ and $\mathbf{q}$-point grids.

The phonons were computed on a $3\times3\times3$ $\Gamma$-centered $\mathbf{q}$-point grid, which corresponds to 4 $\mathbf{q}$-points in the Irreducible Brillouin Zone (IBZ). Since time-reversal symmetry was not implemented in \texttt{EPW} 3, the previous release required 5 $\mathbf{q}$-points instead. This difference in the number of phonon calculations explains the speedup gained in the \texttt{Quantum ESPRESSO} phonon part in Fig.~\ref{Speedup}. 

At the \texttt{EPW} level we split the computational time into three steps: (i) the construction of the electron-phonon matrix elements on the full BZ starting from the variation of the self-consistent potential, which is read from file on the coarse IBZ; (ii) the Wannierization from the  $\mathbf{k}$- and $\mathbf{q}$-point coarse grids to the maximally localized Wannier representation; (iii) the Wannier interpolation from real space to fine $10\times10\times10$ $\mathbf{k}$ and $\mathbf{q}$-point grids. 
These three steps are also possible restart points of the \texttt{EPW} software. 
For this specific case we obtained a $\sim$10 times speedup on the \texttt{EPW} part when the code is run in sequential, as compared to the previous version of the code. 
The test was performed on an Intel Xeon CPU E5-2620 with a clock frequency of 2.00~GHz.
The codes were compiled using ifort 15.0.3 with the following compilation flags \texttt{-O2 -assume byterecl -g -traceback -nomodule -fpp}.

At this stage, the only two levels of parallelization available in \texttt{EPW} are $\mathbf{k}$- and $\mathbf{q}$-point parallelization through message passing interface (MPI). $\mathbf{G}$-vector parallelization is planned for a future release.
\begin{figure}[b!]
  \centering
  \includegraphics[width=0.99\linewidth]{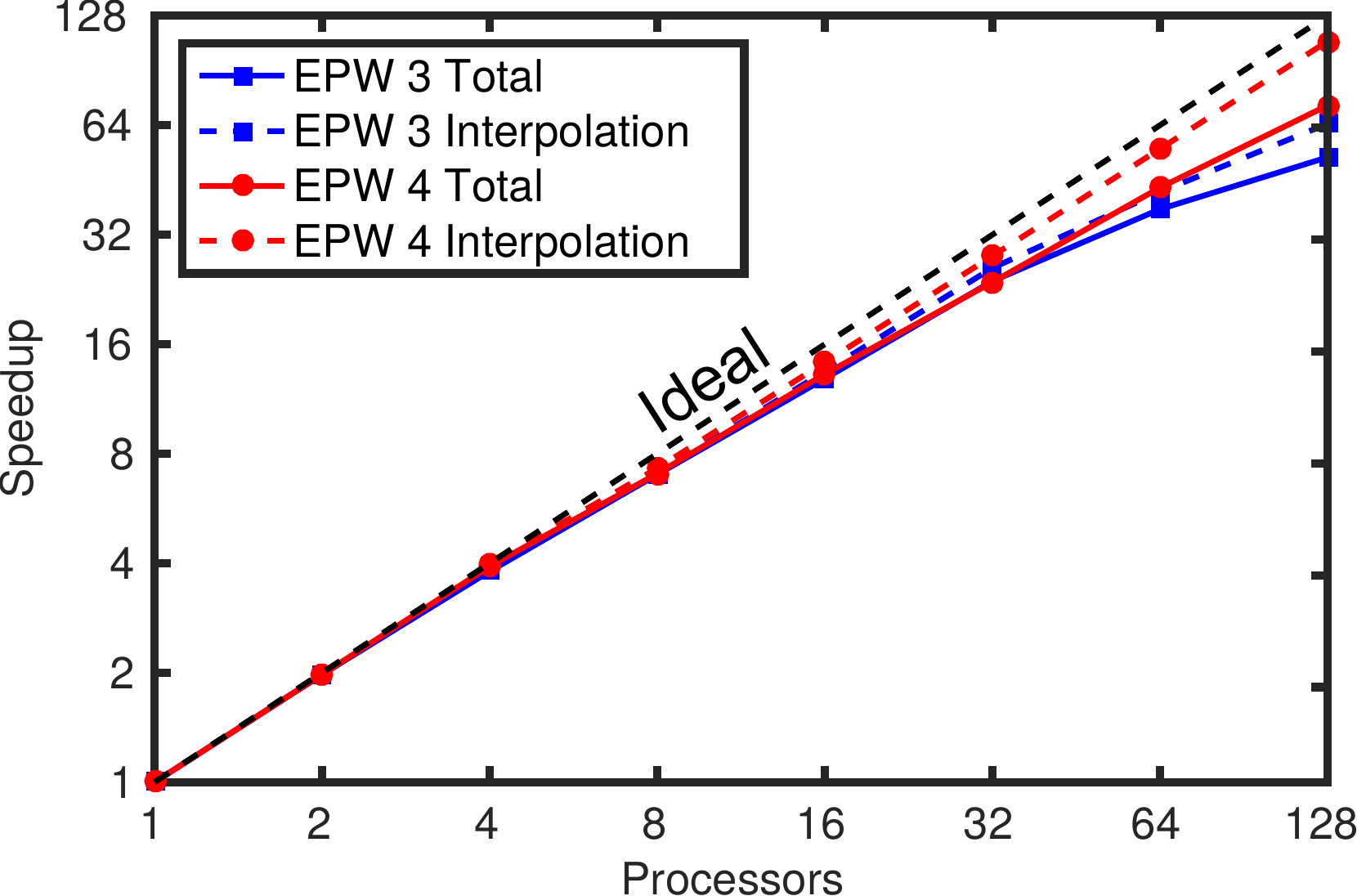}
  \caption{\label{Scalability}(Color online) Parallelization in \texttt{EPW} to compute the electronic lifetime of SiC with the previous and the current version of \texttt{EPW}.  
The blue and red plain lines show the speedup obtained on a full calculation with the previous and current version of \texttt{EPW}. The speedup with 128 processors is 55 and 76 for the previous and current version, respectively.   
The interpolation algorithm (the most time consuming part) has been improved (dashed lines).}
\end{figure}

To test the scalability of the code, we studied SiC using a $6\times6\times6$ $\Gamma$-centered $\mathbf{k}$ and $\mathbf{q}$-points coarse grids. The fine grids on which the Wannier interpolation was performed were a $50\times50\times50$ $\mathbf{k}$-point grid and a $10\times10\times10$ $\mathbf{q}$-point grid.  The test was performed on an Intel Xeon CPU E5620 with a clock frequency of 2.40~GHz. The codes were compiled using ifort 13.0.1 with the following compilation flags \texttt{-O2 -assume byterecl -g -traceback -nomodule -fpp}. The MPI parallelization was performed using Open MPI 1.8.1.

We purposely chose a denser $\mathbf{k}$-point grid to better show the effort placed on the $\mathbf{k}$-point parallelization.
The blue and red solid lines on Fig.~\ref{Scalability} show the speedup obtained in a complete calculation with the previous and current versions of \texttt{EPW}. The speedup with 128 processors is 43\% and 59\% of the ideal speedup for the previous and current version, respectively. 
As shown in Fig.~\ref{Speedup}, the most time-consuming part is the interpolation from real space onto the fine  $\mathbf{k}$ and $\mathbf{q}$-point grids. Therefore specific effort has been dedicated to the scaling of that part of the code. This can be seen on the curve with dashed lines in Fig.~\ref{Scalability}; here the speedup with 128 processors is 51\% and 88\% for the previous and current version, respectively.

\section{Buildbot-based test farm}\label{testfarm}

\begin{table*}[t!]
 \begin{small}
 \begin{center}
   \begin{tabular}{c l l l l l }
   \hline
   \textbf{Slaves} &  \textbf{OS} & \textbf{Compiler} & \textbf{Flags} & \textbf{Parallel} & \textbf{Library} \\
   \hline  
\rowcolor{Gray}    
   farmer1 & Ubuntu 14.04.3  & gcc 4.8.5      & -O3 -g -x f95-cpp-input & -             & internal \\
   farmer1 & Ubuntu 14.04.3  & gcc 4.8.5      & -O3 -g -x f95-cpp-input & openmpi 1.8.8 & internal \\
\rowcolor{Gray}    
   farmer1 & Ubuntu 14.04.3  & intel 12.1.2   & -O2 -assume byterecl -g -fpp  & openmpi 1.8.8	 & MKL \\
   farmer1 & Ubuntu 14.04.3  & intel 13.1.3   & -O3 -xHost -assume byterecl  & openmpi 1.8.8	 & MKL + fftw 3.3.4 \\
           &                 &                & -assume  buffered\_io -fpp   &                 &                  \\
\rowcolor{Gray} 
   farmer1 & Ubuntu 14.04.3  & intel 15.0.3   & -O2 -assume byterecl -g -fpp	& openmpi 1.8.8	 & MKL + fftw 3.3.4 \\
   farmer2 & CentOS 7.1.1503 & gcc 5.2.0      & -O3 -g -x f95-cpp-input & openmpi 1.8.8 &  internal \\
\rowcolor{Gray} 
   farmer2 & CentOS 7.1.1503 & pgf90 15.7     & -fast & mpich 3.1.3 &  fftw 3.3.4 \\
   farmer2 & CentOS 7.1.1503 & intel 13.0.1   & -O3 -xHost -assume byterecl & mpich 3.1.4 & MKL + fftw 3.3.4 \\
           &                 &                & -assume  buffered\_io -fpp   &                 &                  \\
\rowcolor{Gray} 
   farmer2 & CentOS 7.1.1503 & intel 15.0.3   & -O3 -g -x f95-cpp-input & Intel mpi 5.1 & MKL + fftw 3.3.4 \\
   farmer2 & CentOS 7.1.1503 & Nag 6.0        & -O0 -kind=byte -dcfuns -mismatch & - & internal \\
\rowcolor{Gray} 
   farmer3 & openSUSE 13.2   & gcc 4.9.3      & -O3 -g -x f95-cpp-input & mpich 3.1.3 & fftw 3.3.4 \\
   farmer3 & openSUSE 13.2   & intel 15.0.3   & -O3 -xHost -assume byterecl & mvapich2 2.2a & MKL + FFT MKL \\
           &                 &                & -assume  buffered\_io -fpp   &                 &                  \\
\rowcolor{Gray} 
   farmer4 & SLinux 7.1.1503 & intel 15.0.3   & -O2 -assume byterecl -g -fpp &  mvapich2 2.2a  &   MKL + FFT MKL    \\           
   \hline
   \end{tabular}
   \caption{ \label{tabtestfarm} The EPW test-farm is maintained using the Buildbot software. The hardware is as follows: farmer1 and farmer2 are Intel Xeon X5345@2.33Ghz with 16 Gb of RAM; farmer3 is a AMD Opteron P265@1.8Ghz with 2Gb of RAM; farmer4 is an Intel Core 2 Duo@2.8Ghz desktop computer with 4~Gb of RAM.}
 \end{center}
 \end{small}
\end{table*}

The \texttt{EPW} software is maintained and tested using the continuous integration framework Buildbot (\texttt{buildbot.net}). The test farm and waterfall can be found at \texttt{epw.org.uk/Main/TestFarm}. 

The tested operating systems (OS), compilers (with compiler flags), parallelization options, and optimization libraries are given in Tab.~\ref{tabtestfarm}.
The test farm currently operates on four physical hardwares with the following specifications: farmer1 and farmer2 are Intel Xeon X5345@2.33Ghz with 16~Gb of RAM, farmer3 is a AMD Opteron P265@1.8Ghz with 2~Gb of RAM and farmer4 is an Intel Core 2 Duo@2.8Ghz desktop computer with 4~Gb of RAM.  

The Ubuntu, CentOS, openSUSE and Scientific Linux OS distributions are tested in conjunction with the gcc, intel, PGI and NAG compilers. The following MPI implementations are tested: openmpi, mpich, intel mpi and mvapich. The use of internal, MKL and FFTW libraries are also checked.  
Notably, the \texttt{EPW} code has been tested and debugged using the Fortran NAG compiler; this greatly improved the stability of the code. 

The test-suite is currently composed of 10 tests that cover 93\% of the routines and 61\% of the 25,600 software blocks (blocks are defined by 'IF' conditions). Incomplete coverage is in part due to features being in development. We plan to add more tests to improve code coverage. Up to date code coverage information can be found on the \texttt{EPW} website.   

As a result of the above, the code has been cleaned, memory leaks have been corrected, unused and deprecated variables and functions have been cleaned.  
Coding rules and good practice have been enforced within the \texttt{EPW} project.

\section{Examples}
\label{examplessection}

In this section we present some representative physical examples of the capabilities of \texttt{EPW}. Most examples are available as tutorials on the \texttt{EPW} website and are also distributed with the software. 

\subsection{Spectral functions and linewidths of B-doped diamond}

The first example focuses on studying a heavily-doped semiconductor, B-doped diamond. The tutorial can be found in the \texttt{EPW/examples/diamond} folder of \texttt{EPW}. An online version is available at \texttt{epw.org.uk}. 

\begin{figure}[b!]
  \centering
  \includegraphics[width=0.95\linewidth]{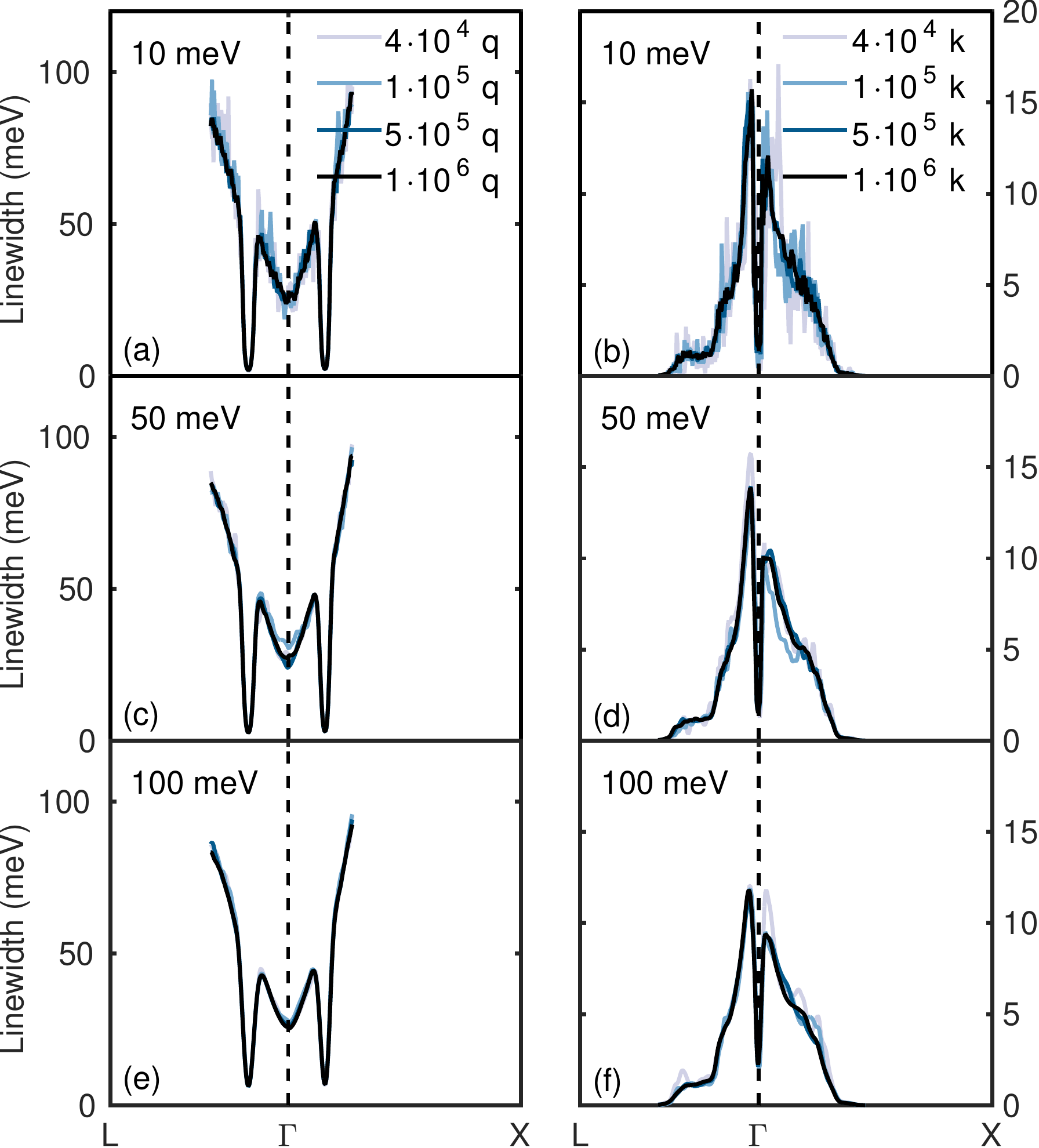}
  \caption{\label{C-linewidths}(Color online) Calculated electron (a),(c),(e) and phonon (b),(d),(f) linewidths at 300~K for the valence band top and for the highest optical mode of B-doped diamond, respectively. The plots are shown for increasing numbers of random \textbf{k}- or \textbf{q}-points used in the BZ integration. The results are also presented for three different broadening parameters of 10, 50 and 100~meV.}
\end{figure}

This tutorial follows the study on B-doped diamond made in Ref.~\cite{Giustino2007} with a B content of 1.85~\%, close to the experimental value~\cite{Ekimov2004}. Such $p$-doping shifts the Fermi level by 0.57~eV below the valence band maximum. 
The calculations are performed using DFT in the LDA and lead to a relaxed lattice parameter of 6.6425~bohr, 1.7~\% below the experimental value~\cite{Ekimov2004}. An energy cutoff of 60~Ry is required to accurately describe the plane-wave basis set ($<$~10 meV/atom). The virtual atom pseudopotential $\text{B}_{x}\text{C}_{1-x}$ with $x=0.0185$ was created, with an ionic charge of $Z=3x+4(1-x)$. 
Fractional occupations resulting from the $p$-doping are described using a Methfessel-Paxton first-order smearing~\cite{Methfessel1989} of 0.02~Ry.
\begin{figure*}[t!]
  \centering
  \includegraphics[width=0.85\linewidth]{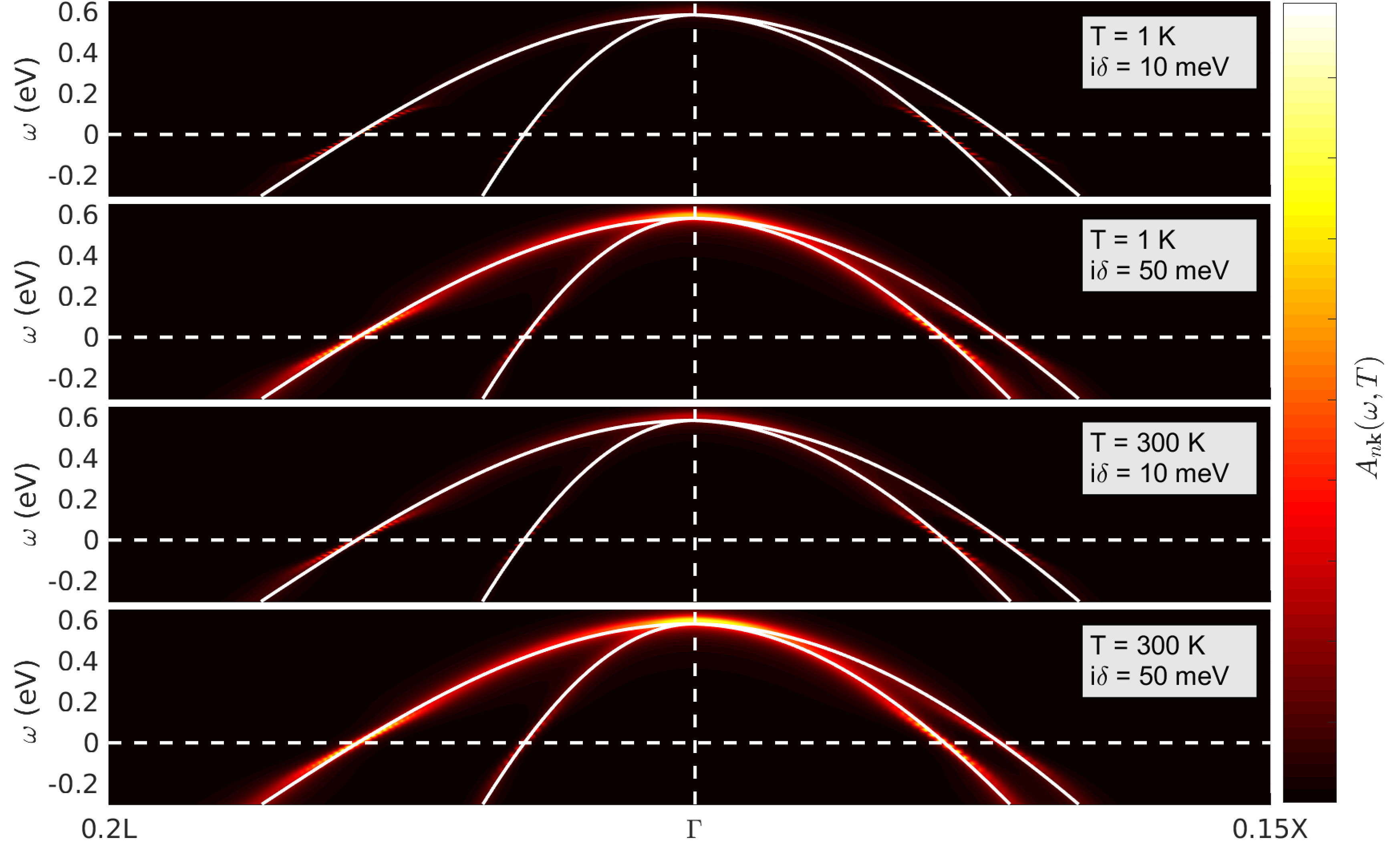}
  \caption{\label{spectralfig}(Color online) Electronic spectral function $A_{n\mathbf{k}}(\omega, T )$ for the valence band of B-doped diamond, calculated using a $10^6$ random $\mathbf{q}$-grid at different temperatures and broadening parameters $\delta$.}
\end{figure*}

We used a $\Gamma$-centered $6\times6\times6$ coarse grid to sample both the electron and phonon-momenta space. Taking advantage of symmetries, we only had to perform 16 DFPT calculations in the irreducible wedge. Such coarse grids are dense enough to perform a very accurate Wannier interpolation~\cite{Giustino2007}.
Using the aforementioned parameters and pseudopotential, the electronic and vibrational bandstructure can be found in Figs.~3 and 4 of Ref.~\cite{Giustino2007}, respectively.  

The electron $\Sigma_{n\mathbf{k}}''(\varepsilon_{n\mathbf{k}})$ and phonon $\Pi_{\mathbf{q}\nu}''(\varepsilon_{n\mathbf{k}})$ linewidths at 300~K  are shown in Fig.~\ref{C-linewidths} for various BZ samplings and for different Gaussian broadenings of the Dirac delta functions of Eqs.~\eqref{linewidthel} and \eqref{linewidthph}. From Fig.~\ref{C-linewidths} we can conclude that $10^6$ random \textbf{k}/\textbf{q}-points are enough to accurately  describe numerically the integrals in Eqs.~\eqref{linewidthel} and \eqref{linewidthph}.
   
It is worth noting that the calculated electron linewidths for the valence band of B-doped diamond shown on the left side of Fig.~\ref{C-linewidths} differ from the ones computed in Ref.~\cite{Giustino2007} where the calculations of electron linewidths were performed for the undoped system.

The electron linewidths show two distinct features: (i) the linewidth has an overall parabolic shape and (ii) there are two dips at the Fermi wavevector $k_{\rm F}$. 
The parabolic shape is due to the fact that, if we approximate the electron-phonon matrix elements as constant, the imaginary part of the self-energy Eq.~\eqref{linewidthel} is directly proportional to the electronic density of states after neglecting the phonon frequency in the delta function. The dips are due to the fact that the self-energy must vanish at the Fermi level, owing to the energy-conservation selection rule.

The phonon linewidths have two distinct features. First, the linewidth of states with momenta larger that the Fermi-surface diameter ($|\mathbf{q}|> 2k_{\rm F}$) is negligible because these phonons cannot generate an electron-hole pair.
This is linked to the shape of the electronic bandstructure and Fermi-level position (see Fig.~3 of Ref.~\cite{Giustino2007}). Second, the linewidth goes to zero (infinite lifetime) at the zone-center. As can be seen in Eq.~\eqref{linewidthph}, this is because the transition is blocked by selection rules when the energy separation between the initial and final state with same momentum \textbf{k} exceeds the largest phonon frequency~\cite{Giustino2007a,Calandra2005}, as is the case in the present example.

Figure~\ref{spectralfig} shows the electronic spectral function of the valence band of B-doped diamond, integrated on a random \textbf{q}-grid containing $10^6$ points, for various temperatures and broadening parameters $\delta$, see Eq.~\eqref{spectral_funct}. 

Most notably, one can see that the lifetime close to the Fermi level is large and that there is the emergence of kinks in the dispersion relations, at energies around the highest phonon energy, on both sides of the Fermi level. The energy renormalization (real part of the electronic self-energy) changes sign across the Fermi level (negative renormalization for $\omega < \omega_{\mathbf{q}\nu}$ and positive otherwise).
This is due to the fact that, in Eq.~\eqref{elselfenergy}, when $\omega <  \omega_{\mathbf{q}}$ the sign of the final state $\varepsilon_{m\mathbf{k+q}}$ gives the overall sign of the self-energy, whereas when $\omega \gg  \omega_{\mathbf{q}}$ then the denominator of Eq.~\eqref{elselfenergy} is positive. 


\subsection{Scattering rate of undoped Si}

We choose to study the scattering rate of undoped silicon, as this material has been studied previously by several groups using parametrized tight-binding models~\cite{Rideau2011}, direct evaluations of the electron-phonon matrix elements~\cite{Restrepo2009,Sun2012}, \textit{ab-initio} linear interpolations~\cite{Li2015} as well as Wannier interpolation using previous versions of the \texttt{EPW} code~\cite{Bernardi2014,Qiu2015,Tandon2015}.
\begin{figure}[b!]
  \centering
  \includegraphics[width=0.95\linewidth]{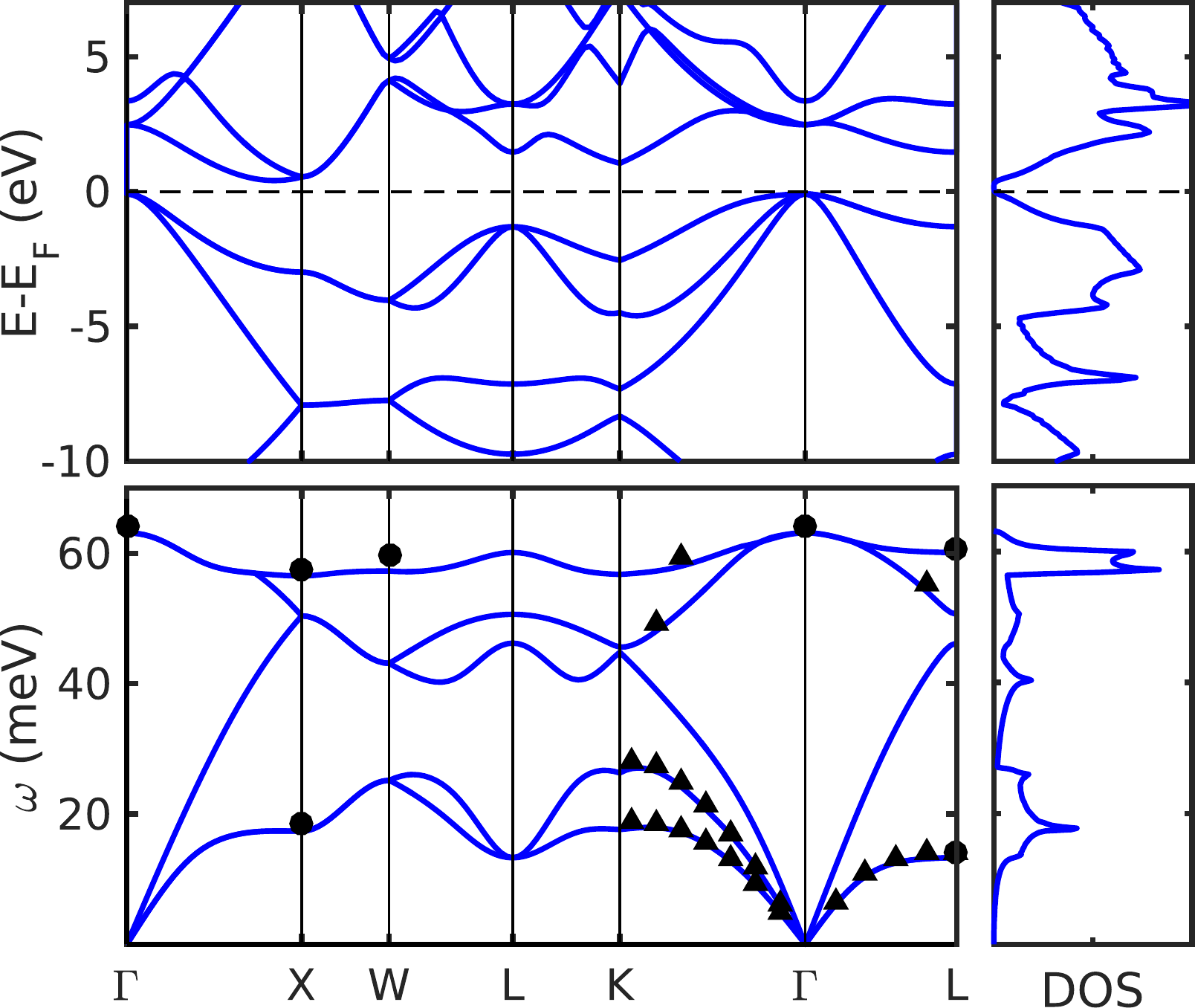}
  \caption{\label{BS-Si} (Color online) Electron (top) and phonon (bottom) bandstructures and density of states (DOS) of Si at the optimized lattice parameter. Experimental neutron-scattering data measured at 300~K are taken from Ref.~\cite{Dolling1963} (filled circles) and Ref.~\cite{Nilsson1972} (filled triangles). }
\end{figure}
We recover the same results, including the sharp increase in scattering rate above the optical phonon emission threshold starting from very coarse initial grids.

The Si norm-conserving pseudopotential used for this study was generated from a non-relativistic calculation. The valence electrons treated explicitly in the calculations are the 3s$^{2}$3p$^2$ with the Perdew and Zunger parametrization of the LDA~\cite{Perdew1981}.
Convergence studies (with errors below 2~meV/atom on the total energy) lead to using an $8\times8\times8$~$\Gamma$-centered Monkhorst-Pack~\cite{Monkhorst1976} $\mathbf{k}$-point sampling of the BZ, and a plane wave energy cutoff of 45~Ry.

The lattice parameter obtained after structural relaxation is calculated to be 10.207~bohr, slightly below the experimental lattice parameter of 10.262~bohr~\cite{OMara1990}. 
The electronic bandstructure shown at the top of Fig.~\ref{BS-Si} was computed at the DFT level and gave an indirect bandgap of 0.497~eV,  underestimating as expected the experimental value of 1.12~eV.
The calculated phonon bandstructure shown at the bottom of Fig.~\ref{BS-Si} slightly underestimates the neutron scattering data measured at 300~K~\cite{Dolling1963, Nilsson1972}.

\begin{figure}[t!]
  \centering
  \includegraphics[width=0.99\linewidth]{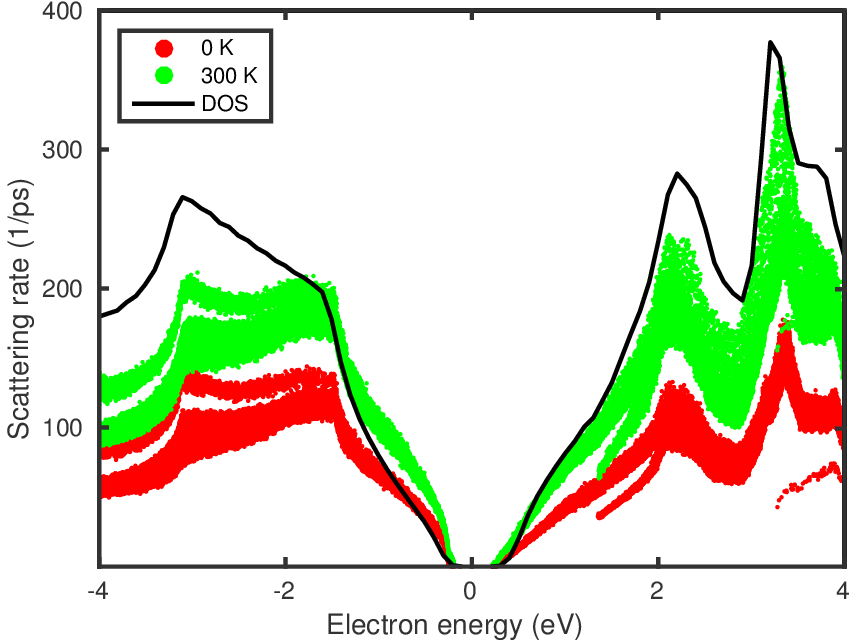}
  \caption{\label{scatering-Si} (Color online) Carrier scattering rate in Si as a function of energy at 0~K (red dots) and 300~K (green dots), obtained with a broadening of 10~meV and 30,000 random $\mathbf{k}$-points, 150,000 random $\mathbf{q}$-points. The electronic DOS is superimposed with arbitrary units. The Fermi level is placed in the middle of the gap.}
\end{figure}

\begin{figure}[t!]
  \centering
  \includegraphics[width=0.99\linewidth]{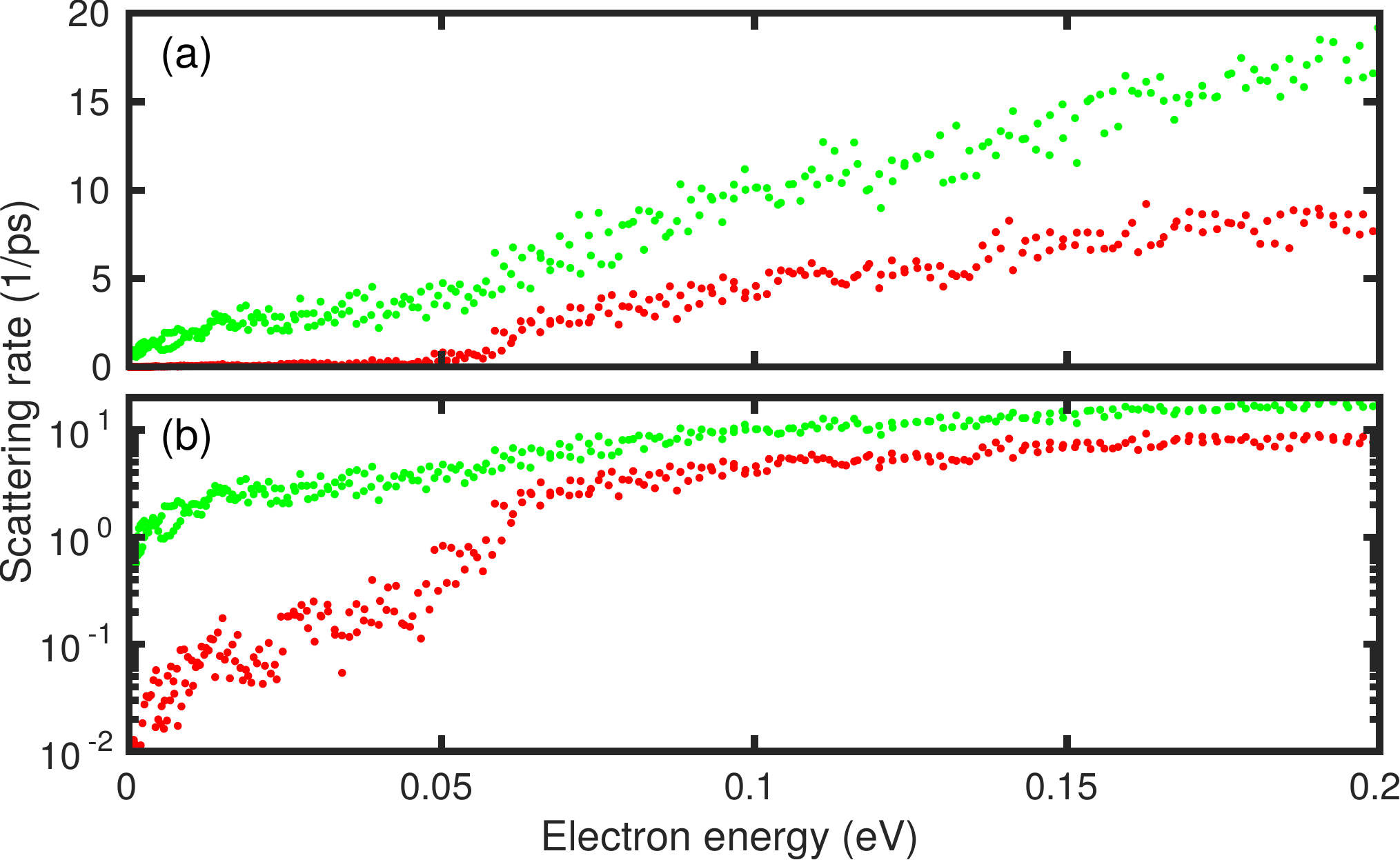}
  \caption{\label{scatering-Si2} (Color online) Carrier scattering rates in Si as a function of energy with respect to the conduction band minimum, at 0~K (red dots) and 300~K (green dots). The data were obtained with a broadening of 1~meV, 500 $\mathbf{k}$-points along the $0.25\Delta-X$ line, and 500,000 random $\mathbf{q}$-points. Both images show the same data on a linear (a) and semi-logarithmic (b) scale. A clear increase in the scattering rate appears around 60~meV, where the scattering by optical phonons becomes allowed.  }
\end{figure}

The scattering rate is the inverse of the relaxation time $\tau_{n\mathbf{k}}$, and is directly connected to the imaginary part of the electron self-energy from Eq.~\eqref{linewidthel}:
\begin{equation}\label{scatteringRate}
\frac{1}{\tau_{n\mathbf{k}}(\omega,T)} = 2 \Sigma_{n\mathbf{k}}^{''}(\omega,T).
\end{equation}

This quantity is very sensitive to the details of the electronic structure and therefore requires dense interpolated $\mathbf{k}$-point and $\mathbf{q}$-point grids. Nevertheless, owing to the accuracy of Wannier interpolation, we only need relatively coarse initial grids. For this study we used $6\times6\times6$ coarse $\mathbf{k}$-point and $\mathbf{q}$-point grids (16 $\mathbf{q}$-points in the IBZ). 
Fig.~\ref{scatering-Si} shows the scattering rate as a function of energy at 0~K and 300~K, obtained with a broadening of 10~meV and 30,000 random $\mathbf{k}$-points, 150,000 random $\mathbf{q}$-points. The imaginary part of the self-energy follows closely the electronic density of state (DOS). This is expected since the accessible phase space reflects the DOS. Raising the temperature increases the scattering rate almost uniformly.

It is important to use a sufficiently small Gaussian broadening to avoid washing out detailed features of the scattering rate. For this reason, we also present in Fig.~\ref{scatering-Si2} a zoom of the scattering rate along the high symmetry $0.25\Delta-X$ line, using Gaussian broadening of 1~meV.     
The scattering rate is small at zero temperature for energies lower than the optical phonons (60~meV), since only low-energy phonons can be emitted. 
The evaluation of the scattering rate is a key quantity to study other phenomena such as carrier mobilities. However, here we want to emphasize that such calculations are computationally affordable, and can now be carried out on a standard desktop computer.

\subsection{Spectral function and electronic resistivity of Pb with and without spin-orbit coupling}

The third example presents a study on metallic Pb, focusing on the effect of spin-orbit coupling (SOC). This tutorial can be found in the \texttt{EPW/examples/pb} folder of \texttt{EPW}. An online version is available at \texttt{epw.org.uk}. 
We reproduce the study by Heid and co-workers~\cite{Heid2010}, at a significantly lower computational cost owing to Wannier interpolation. 

\begin{figure}[t!]
  \centering
  \includegraphics[width=0.99\linewidth]{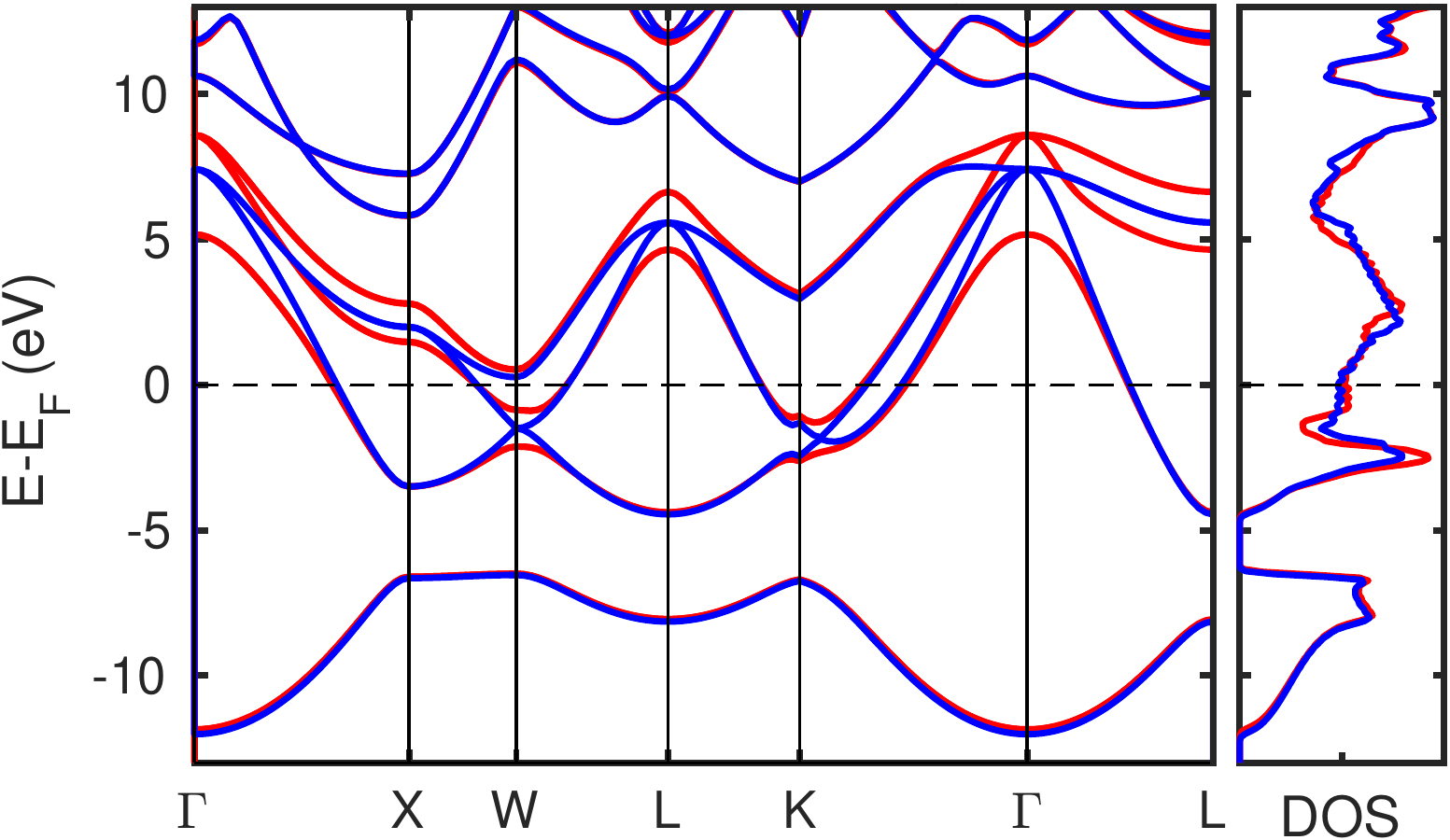}
  \caption{\label{BS-Pb} (Color online) Electronic bandstructure and DOS of Pb at the optimized lattice parameter with (red line) and without (blue line) spin-orbit interaction.}
\end{figure}
\begin{figure}[t!]
  \centering
  \includegraphics[width=0.99\linewidth]{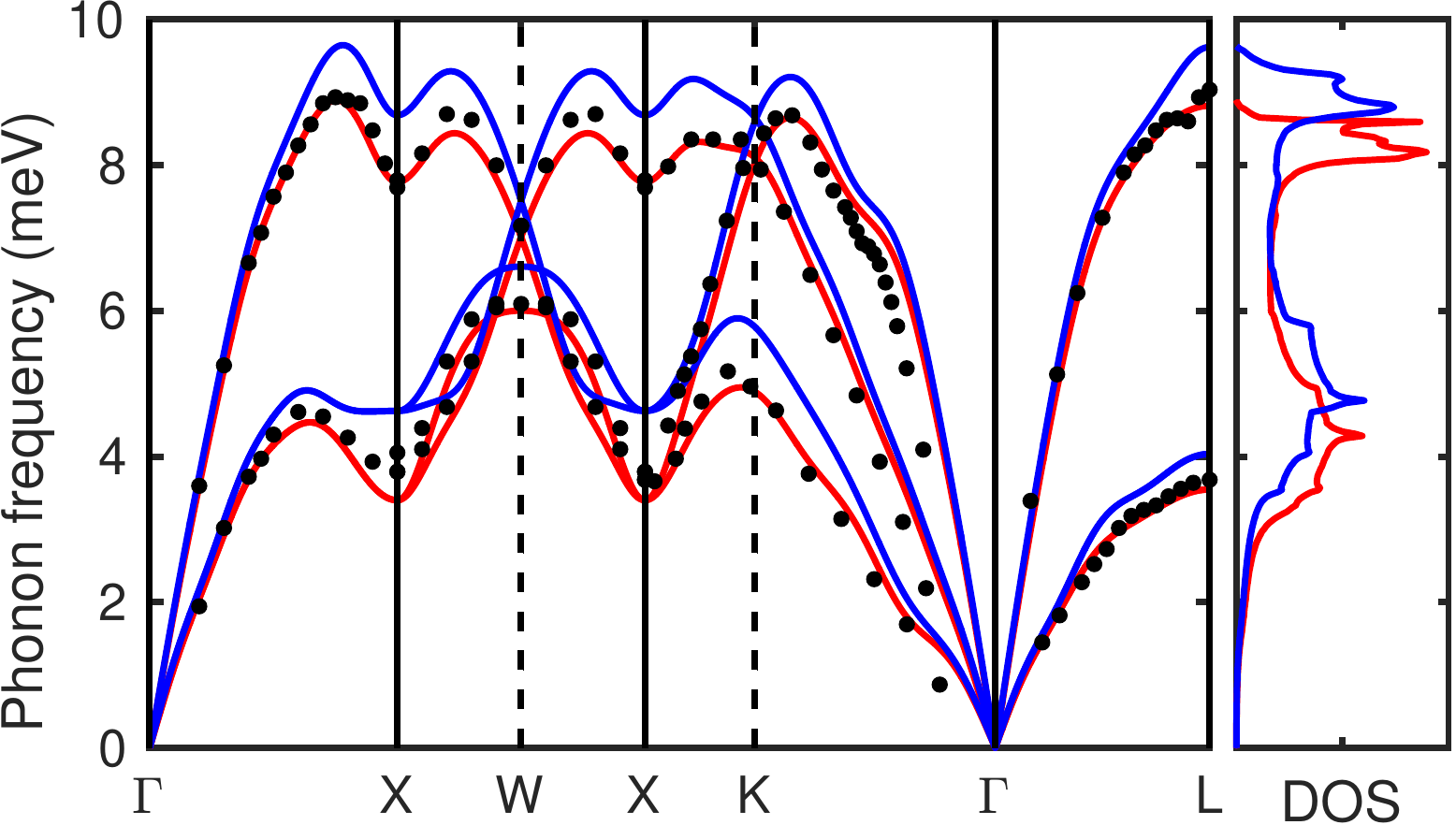}
  \caption{\label{PH-Pb} (Color online) Phonon dispersion relations and phonon density of states (PDOS) of Pb at the optimized lattice parameter with (red line) and without (blue line) spin-orbit interaction. The experimental data (black dots) at 100~K from neutron scattering are taken from Ref.~\cite{Brockhouse1962}.}
\end{figure}

For this study, we used a norm-conserving semi-core fully relativistic pseudopotential where the $5d^{10}6s^{2}6p^{2}$ electrons are considered as valence electrons. Structural relaxation has been performed using DFT within the LDA~\cite{Ceperley1980,Perdew1981} and a planewave basis set as implemented in the \texttt{Quantum ESPRESSO}~\cite{Giannozzi2009} code. A planewave energy cutoff of 90~Ry and a $8\times8\times8$ shifted \textbf{k}-point mesh were required to converge the total energy below 0.5~mRy. A Gaussian smearing of 0.05~Ry was used in the BZ integration.
Using these parameters, we performed a structural relaxation on the face-centered cubic Pb and obtained a lattice parameter of 9.222~bohr without SOC, and 9.270~bohr when the SOC was included. The latter value agrees well with the low-temperature (5~K) experimental lattice parameter of 9.269~bohr~\cite{Touloukian1975,Grabowski2007}.    

\begin{figure*}[t!]
  \centering
  \includegraphics[width=0.8\linewidth]{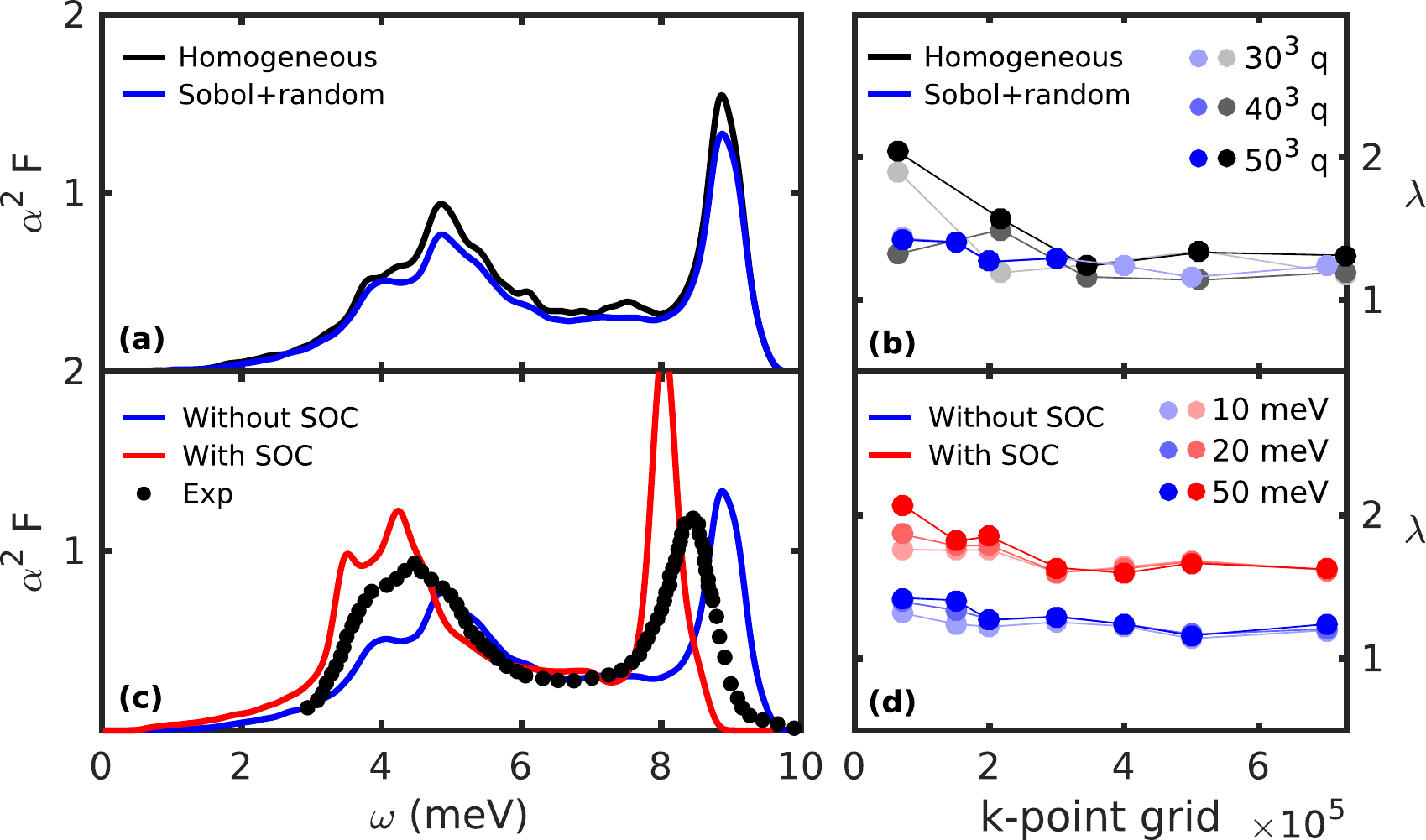}
  \caption{\label{a2F-Pb} (Color online) 
   Effect of different fine sampling methods in \texttt{EPW} (homogeneous $\mathbf{k}$ and $\mathbf{q}$-point integration against quasi random $\mathbf{q}$-point using a Sobol algorithm) on (a) the Eliashberg $\alpha^2F$ and (b) the electron-phonon coupling strength $\lambda$ of Pb without SOC. Effect of the SOC on (c) the Eliashberg $\alpha^2F$ with an electron smearing of 10~meV and (d) on $\lambda$ when different electron smearing are considered. The phonon smearing is 0.15~meV. The experimental data points are taken from Ref.~\cite{Scalapino1969}. }
\end{figure*}

The associated Kohn-Sham electronic bandstructure is presented in Fig.~\ref{BS-Pb} along the main high-symmetry lines of the BZ. We see that SOC leads to avoided crossings near the Fermi level, but the Fermi surface itself is almost unmodified~\cite{Corso2008}.

Due to the complex Fermi surface of Pb, a $14\times14\times14$ $\mathbf{k}$-point and a $10\times10\times10$ $\mathbf{q}$-point grid with a Gaussian broadening of 0.025~Ry were required to achieve convergence. As already shown by Heid and co-workers~\cite{Heid2010}, we find that calculations performed without SOC overestimate the phonon frequencies with respect to experiment, while calculations including SOC are much closer to the neutron scattering data measured at 100~K~\cite{Brockhouse1962} (see Fig.~\ref{PH-Pb}). 

\begin{figure}[b!]
  \centering
  \includegraphics[width=0.99\linewidth]{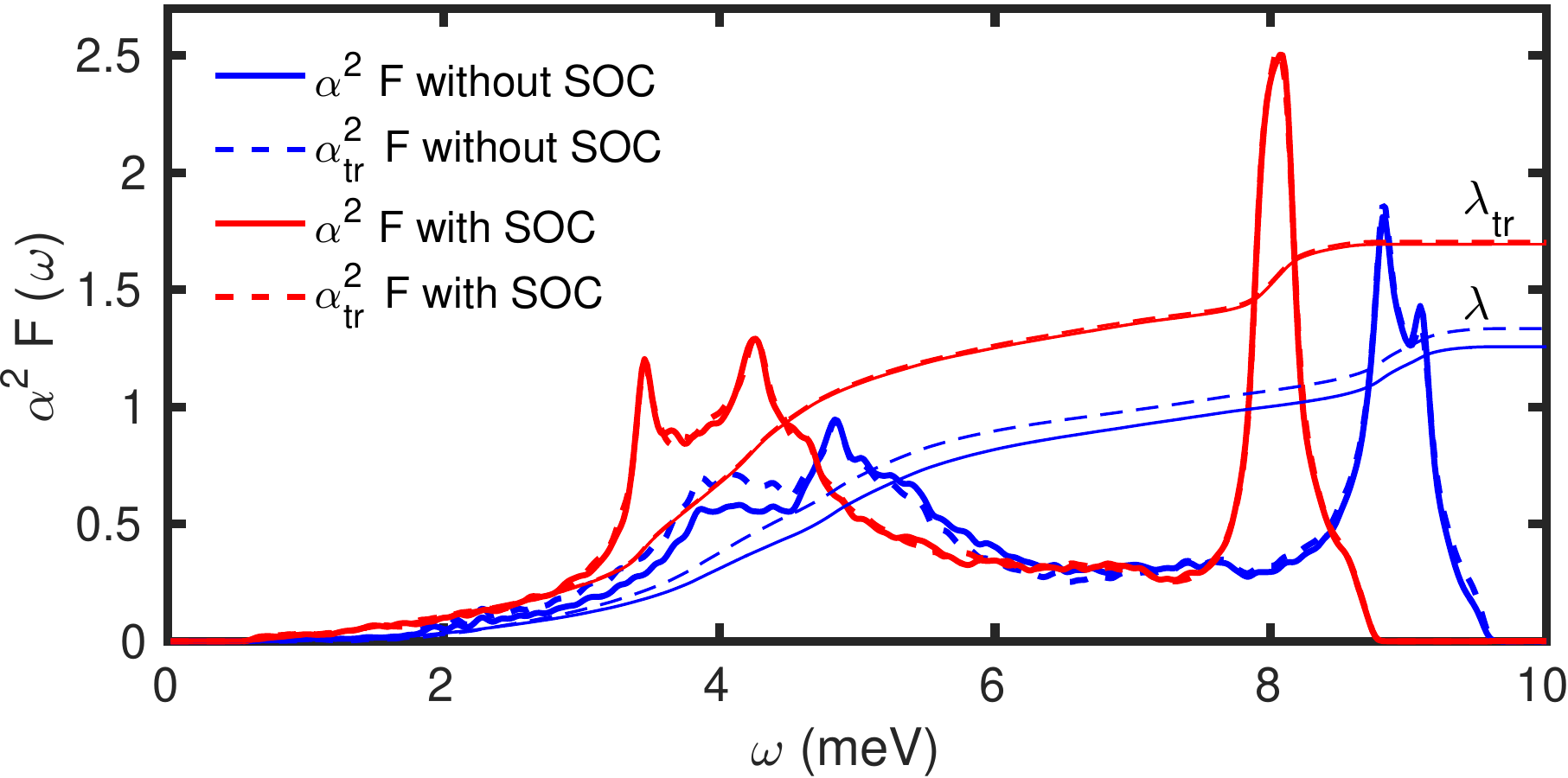}
  \caption{\label{a2F-Pbtr} (Color online) Calculated isotropic Eliashberg spectral function $\alpha^2 F(\omega)$ of Pb (blue solid line) and integrated electron-phonon coupling strength $\lambda$ (thin blue solid line), compared with the Eliashberg transport function (blue dashed line) and the transport coupling strength $\lambda_{\rm tr}$ (thin dashed blue line). The same data are presented with spin-orbit coupling in red. }
\end{figure}

To reduce the computational complexity, we noticed that a $10\times10\times10$ $\mathbf{k}$-point and a $6\times6\times6$ $\mathbf{q}$-point grid
was enough to obtain phonon frequency within 2~cm$^{-1}$ of the grid used in Fig.~\ref{PH-Pb}. 
Once the electron-lattice vibration calculations were performed, we used a $6\times6\times6$ $\mathbf{k}$-point and $\mathbf{q}$-point coarse grids  in \texttt{EPW} for subsequent electron-phonon calculations.    
The interpolated fine grids need to be very dense to correctly sample the complex Fermi surface. 
The Eliashberg spectral function $\alpha^2F$ of Pb was calculated with Eq.~\eqref{isotropiceliash} using a homogeneous and a quasi random 3D Sobol sampling.  
The 3D Sobol quasi random generation of points was performed by skipping the first 1000 values and then retaining every 101st point~\cite{Bratley1988}. An additional random linear scramble combined with a digital shift~\cite{Hong2003} was applied as coded in the software Matlab R2015a.

The total electron-phonon coupling strength $\lambda$ (obtained by averaging Eq.~\eqref{lambdaEQ} over all phonons) for Pb without SOC shows that the quasi random Sobol integration outperforms the standard homogeneous grid sampling (Fig.~\ref{a2F-Pb}b). 
A $\mathbf{k}$-point grid of 400,000 Sobol points with 27,000 random $\mathbf{q}$ points is required to reach convergence. Using those converged values, the Eliashberg spectral function is shown in Fig.~\ref{a2F-Pb}a and reveals that calculations without SOC overestimate the experimental spectral function as expected from the phonon DOS of Fig.~\ref{PH-Pb}. The experimental spectral function was obtained from tunneling experiments at 1~K~\cite{Scalapino1969}. 

Figure~\ref{a2F-Pb}c further shows the effect of including SOC using Sobol sampling with the highest $\mathbf{q}$-point grid used in Fig.~\ref{a2F-Pb}b. Finally, Fig.~\ref{a2F-Pb}d shows that the electron smearing for converged grids does not significantly affect the electron-phonon coupling strength. The phonon smearing used for all the calculations was 0.15 meV.    

After assessing convergence and the effect of SOC, we study basic electronic transport properties. As discussed in Section~\ref{transportProp}, the key quantity is the evaluation of the electron velocity.  
Using the electron velocity in the local approximation we can compute the Eliashberg transport coupling function $\alpha_{\text{tr}}^2F$.  
Fig.~\ref{a2F-Pbtr} shows that 
both the Eliashberg function and its transport counterpart have very similar magnitude and shape in the case of Pb including SOC.
In this case, we find $\lambda_{\text{tr}}=1.69$.
In general $\lambda$ and $\lambda_{\text{tr}}$ are expected to be different~\cite{Grimvall1981}.
These results were obtained using 300,000 quasi random Sobol $\mathbf{k}$-point and 50,000 random $\mathbf{q}$-point meshes, with an electronic Gaussian broadening of 10~meV.

\begin{figure}[t!]
  \centering
  \includegraphics[width=0.99\linewidth]{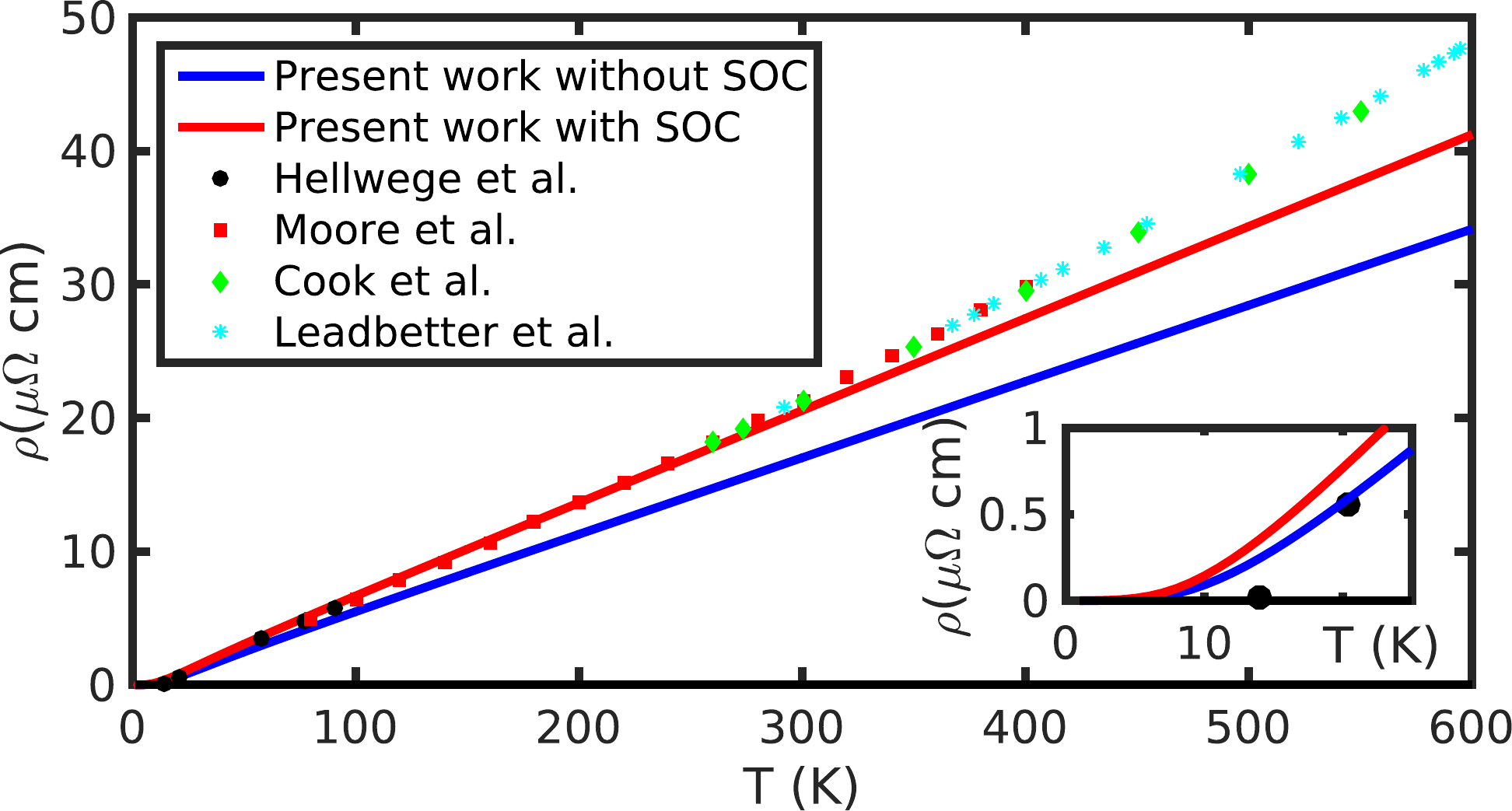}
  \caption{\label{resistivity-Pb} (Color online) Calculated resistivity with (plain red line) and without (plain blue line) spin-orbit coupling. The inset shows a zoom of the resistivity on the 0-25~K temperature range. The experimental measurements are from Refs.~\cite{Leadbetter1966,Moore1973,Cook1974,Hellwege1983}.  }
\end{figure}

Finally, using the Eliashberg transport function, we can evaluate the resistivity with and without spin-orbit coupling using Ziman's resistivity formula, Eq.~\eqref{resistivitytr}. This is shown in Fig.~\ref{resistivity-Pb}. The calculations were performed using 400,000 quasi random Sobol $\mathbf{k}$-points and 27,000 random $\mathbf{q}$-points, with an electronic Gaussian broadening of 20~meV. Our results compare very well with experimental measurements from Refs.~\cite{Leadbetter1966,Moore1973,Cook1974,Hellwege1983}. 
We note that the inclusion of SOC significantly improves the agreement with experiment.

\subsection{Polar coupling in GaN}

\begin{figure}[t!]
  \centering
  \includegraphics[width=0.9\linewidth]{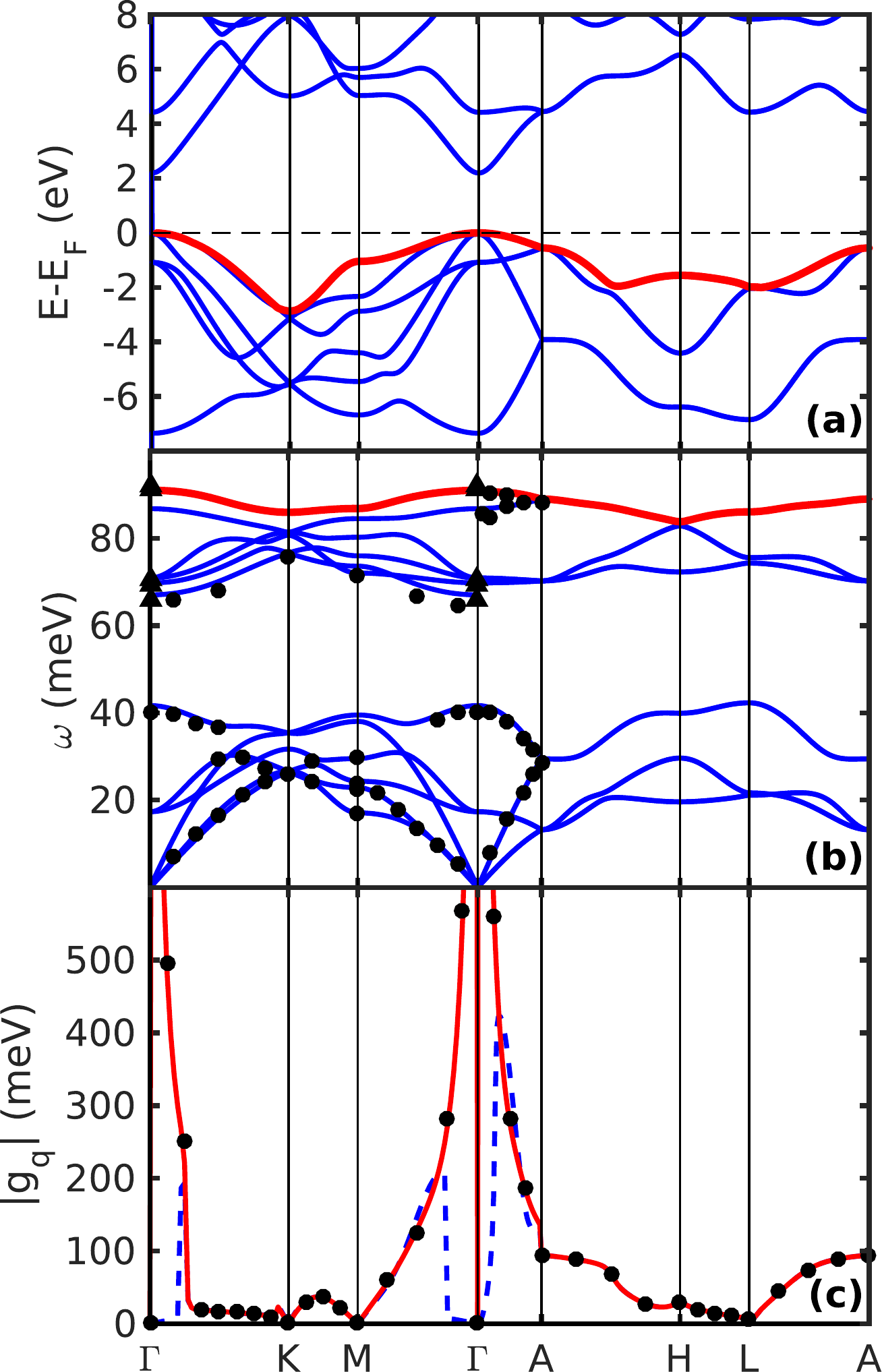}
  \caption{\label{GaN-interpolation} (Color online) (a) Electronic bandstructure of w-GaN along high-symmetry lines in the BZ. The red line highlights the band considered in the bottom panel. (b) Phonon dispersion interpolated from a $6\times6\times6$ $\Gamma$-centered $\mathbf{q}$-point grid. The red line is the optical mode considered in the bottom figure. The filled dots are inelastic X-ray scattering data~\cite{Ruf2001} and the filled triangles are Raman data~\cite{Siegle1997}.  (c) Calculated electron-phonon matrix elements at $\mathbf{k}=\Gamma$ for the highest band and mode index, starting from a $6\times6\times6$ electron and phonon grids. A gauge-invariant electron-phonon matrix element is obtained by averaging the square moduli over degenerate states. The blue dashed line shows the standard interpolation, and the red line shows the polar interpolation using Eq.~\eqref{polar_G}. The two sets of data are compared with direct DFPT calculations at each wavevector (filled dots).   }
\end{figure}

To illustrate the new capability of \texttt{EPW} to treat polar materials, we studied the wurtzite gallium nitride (w-GaN).
For this example, the pseudopotentials were generated using the scalar-relativistic ONCVPSP software~\cite{Hamann2013} and are part of the Schlipf-Gygi norm-conserving pseudopotential library version 1.0~\cite{Schlipf2015}. The valence electrons treated explicitly in the calculations are the  3d$^{10}$4s$^2$4p$^1$  and 2s$^2$2p$^3$  orbitals for Ga and N, respectively.  The Perdew and Zunger parametrization of LDA~\cite{Perdew1981} was used.
Convergence studies lead to the use of a $6\times6\times6$~$\Gamma$-centered Monkhorst-Pack~\cite{Monkhorst1976} $\mathbf{k}$-point sampling of the BZ and a plane wave basis set energy cutoff of 160~Ry (the errors in the total energy are below 3~meV/atom). 

The lattice parameters obtained after structural relaxation were found to be $a = 5.955$~bohr and $c = 9.707$~bohr, 1.1~\% below the experimental lattice parameters~\cite{Bougrov2001}. 
The electronic bandstructure shown on top of Fig.~\ref{GaN-interpolation} was computed at the DFT level, and gave a 2.2~eV direct bandgap at $\Gamma$, underestimating the experimental value of 3.47~eV~\cite{Bougrov2001} as expected.

The calculated phonon bandstructure in the middle of Fig.~\ref{GaN-interpolation} slightly overestimates the single crystal inelastic X-ray scattering data (filled dots)~\cite{Ruf2001} and Raman data (filled triangles)~\cite{Siegle1997}.

We present at the bottom of Fig.~\ref{GaN-interpolation} the direct DFPT calculation of the electron-phonon matrix elements $|g_{mn,\nu}(\mathbf{k,q})|$ along high-symmetry directions of the BZ (filled dots) at $\mathbf{k}=\Gamma$, $m=n$ being the highest valence band, and $\nu$ the highest optical mode. The selected band and mode are indicated in red in the top and middle panels of Fig.~\ref{GaN-interpolation}. The square moduli of electron-phonon matrix elements are averaged over degenerate bands and modes. 
The dashed blue line shows the standard Wannier interpolation of Ref.~\cite{Giustino2007}, and has to be compared with the red line that implements the polar interpolation of Eq.~\eqref{polar_G}~\cite{Verdi2015}. 

Figure~\ref{GaN-interpolation} clearly shows the importance of correctly treating the long-range interaction in the Wannier interpolation when studying polar semiconductors and insulators. This feature is activated in \texttt{EPW} by using the \texttt{lpolar} input variable. 


\subsection{Superconductivity in MgB$_2$}

The last example presented here focuses on the superconducting properties of magnesium diboride (MgB$_2$). This tutorial can be found in the \texttt{EPW/examples/mgb2} folder of \texttt{EPW}. An online version is available at \texttt{epw.org.uk}. 
This tutorial is based on the work of Ref.~\cite{Margine2013} and includes some additional results. 

\begin{figure}[b!]
  \centering
  \includegraphics[width=0.99\linewidth]{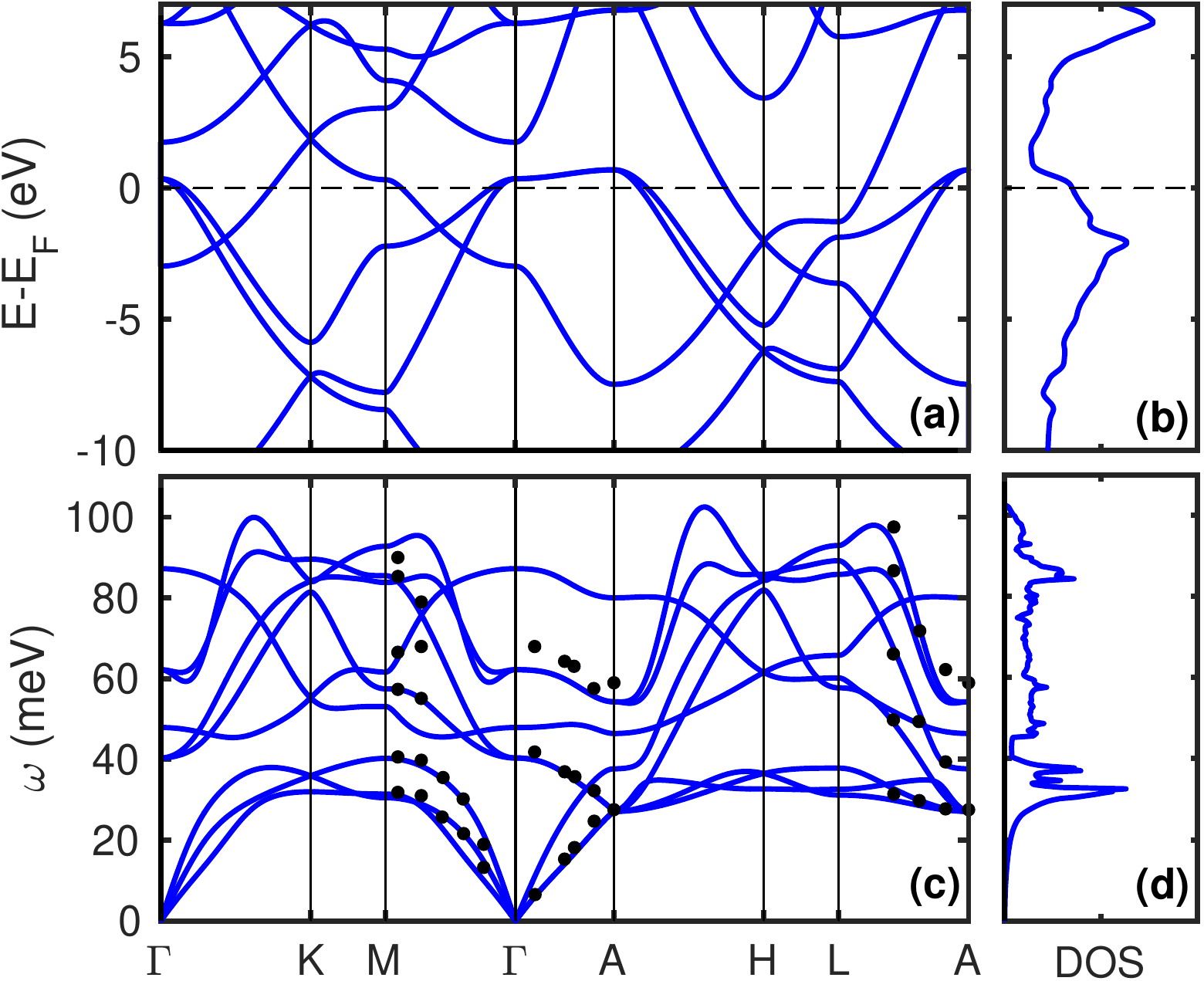}
  \caption{\label{PH-MgB2} (Color online) (a) Electronic bandstructure and (b) DOS of bulk MgB$_2$; (c) phonon bandstructure and (d) PDOS at the experimental lattice parameters. The inelastic X-ray scattering experimental data (black dots) at 300~K are taken from Ref.~\cite{Shukla2003}.  }
\end{figure}

MgB$_2$ is the prototypical multi-gap phonon-mediated superconductor, with a critical temperature of $T_{\rm c} = 39$~K~\cite{Nagamatsu2001}.
The anisotropic gap arises from the $\sigma$ and $\pi$ Fermi-surface sheets~\cite{Kortus2001, Liu2001, Giubileo2001, Choi2002, Choi2002a, Floris2005}.
This superconductor has been widely investigated theoretically~\cite{Hinks2000,Kortus2001,Bouquet2001,Yang2001,Wang2001,Choi2002,Choi2003,Soda2004,Punpocha2004,Dolgov2005,Mishonov2005,Xi2008,Choi2009,Abah2010,Aperis2015} due to its high T$_{\rm c}$.
We re-investigate here MgB$_2$ using \texttt{EPW} to show that its fascinating properties can now be computed easily and accurately. 

\begin{figure}[b!]
  \centering
  \includegraphics[width=0.99\linewidth]{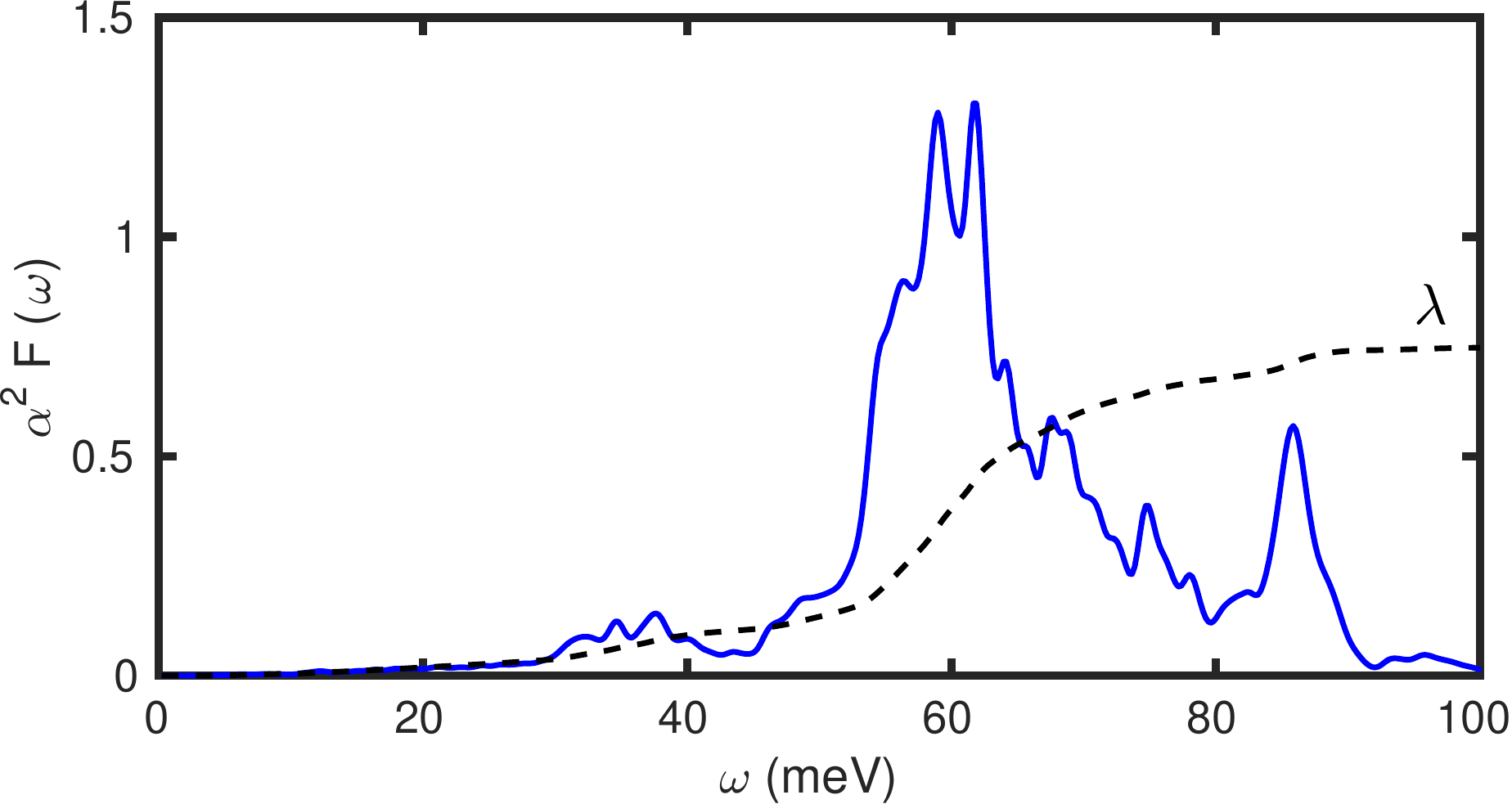}
  \caption{\label{a2F-MgB2} (Color online) Calculated isotropic Eliashberg spectral function $\alpha^2 F(\omega)$ of MgB$_2$ (blue solid line) and integrated electron-phonon coupling strength $\lambda$ (black dashed line).}
\end{figure}

The calculations were performed within DFT-LDA~\cite{Ceperley1980,Perdew1981} using the \texttt{Quantum ESPRESSO}~\cite{Giannozzi2009} code. 
The valence electronic wavefunctions were expanded in a plane-wave basis set using a kinetic energy cutoff of 60~Ry and the charge density is integrated on a $\Gamma$-centered  $24\times24\times24$ \textbf{k}-point mesh. The experimental lattice parameters $a = 5.826$ and $c/a = 1.142$~bohr were used throughout the calculations~\cite{Nagamatsu2001}. 
A Methfessel-Paxton first-order smearing~\cite{Methfessel1989} of 0.02~Ry was applied. 
The first-order potential perturbation and dynamical matrices were calculated using DFPT~\cite{Baroni1987,Gonze1997,Baroni2001} on an irreducible $6\times6\times6$ $\Gamma$-centered \textbf{q}-point mesh.

The electronic bandstructure is shown on the top panel of Fig.~\ref{PH-MgB2}.
The associated phonon bandstructure at the bottom of Fig.~\ref{PH-MgB2} slightly underestimates the inelastic X-ray scattering data measured at 300~K~\cite{Shukla2003}.

\begin{figure}[b!]
  \centering
  \includegraphics[width=0.99\linewidth]{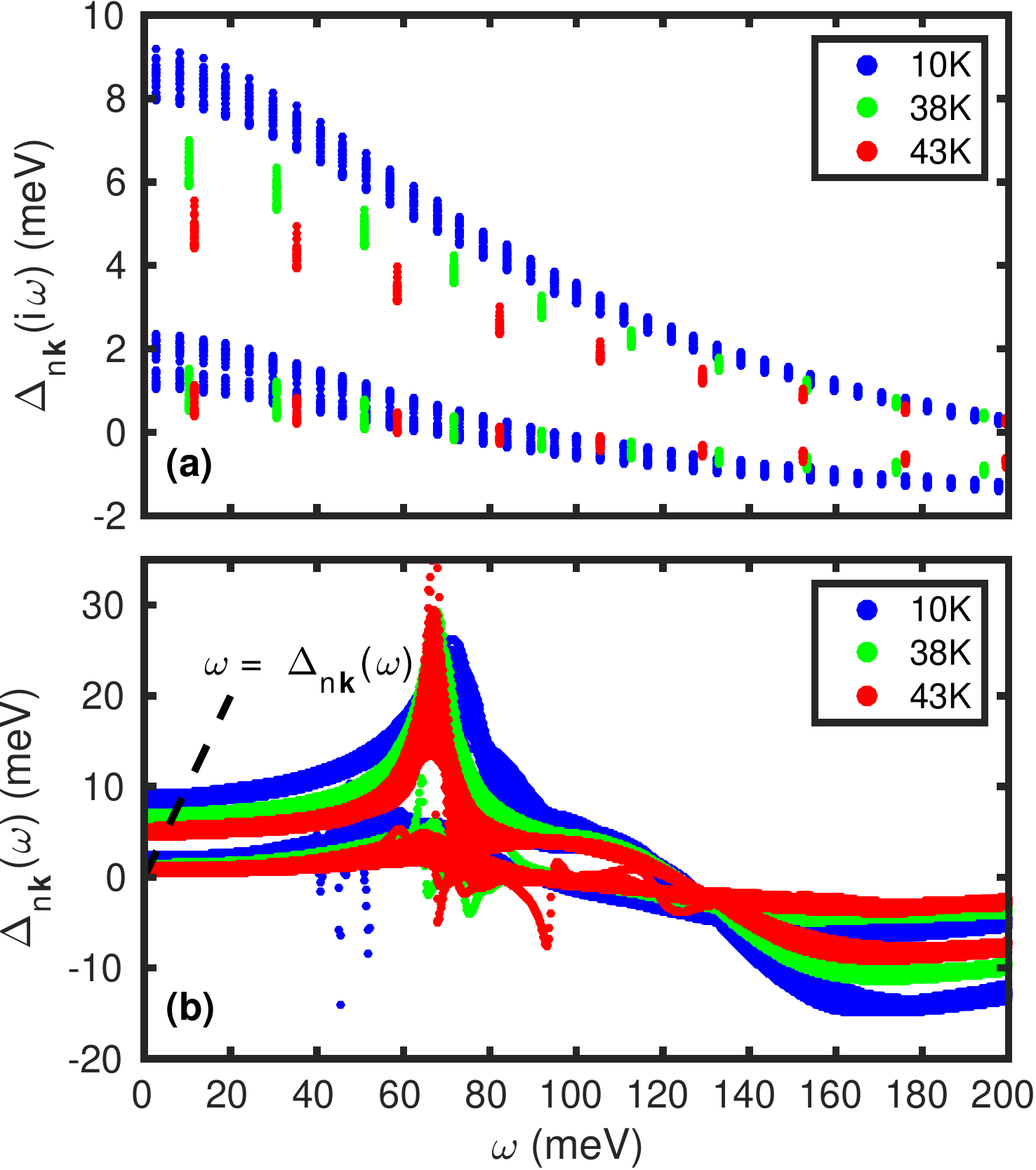}
  \caption{\label{aniso-MgB2} (Color online) Calculated energy-dependent superconducting gap of MgB$_2$ from the anisotropic Eliashberg theory at 10~K (blue), 38~K (green) and 43~K (red) along the imaginary energy axis (a) and along the real energy axis using analytic continuation with Pad\'e functions (b). We only show results for states $n\mathbf{k}$ which have a Kohn-Sham energy 0.1~eV around the Fermi level.  }
\end{figure}

The isotropic Eliashberg spectral function $\alpha^2 F(\omega)$ calculated using Eq.~\eqref{isotropiceliash}, and the associated cumulative electron-phonon coupling strength $\lambda(\omega)$, are shown in Fig.~\ref{a2F-MgB2}.
The spectral function shows a dominant peak around 60~meV, and a secondary one around 86~meV.
The cumulative electron-phonon coupling strength shown as dashed line in Fig.~\ref{a2F-MgB2} is calculated as: 
\begin{equation}
\lambda(\omega) = \int_0^\omega d\omega \frac{2\alpha^2F(\omega)}{\omega}.
\end{equation}

The results are obtained by integrating on a $60\times60\times60$ homogeneous $\mathbf{k}$-point mesh and a $30\times30\times30$ homogeneous $\mathbf{q}$-point mesh with a Gaussian smearing of 100~meV for the electrons, and 0.65~meV for the phonons. The total coupling is $\lambda = 0.75$.
This is in agreement with previous studies where the isotropic electron-phonon coupling strength ranges from 0.71 to 0.78~\cite{Bohnen2001, Choi2002,Floris2005,Eiguren2008,Calandra2010}.

Figure~\ref{aniso-MgB2} presents the calculated energy-dependent superconducting gap of MgB$_2$ from the anisotropic Eliashberg theory.
We show the gap distribution at different temperatures along the imaginary energy axis (top) and along the real energy axis, as obtained by using analytic continuation with Pad\'e functions (bottom). Fig.~\ref{aniso-MgB2} only shows results for states $n\mathbf{k}$ which have a Kohn-Sham energy within 0.1~eV from the Fermi level.

It is worth noticing the red shift of the main peak of $\Delta_{n\mathbf{k}}(\omega)$ with temperature, as well as the flatting of the overall structure towards zero. We note that the solutions on the imaginary frequency axis become more spaced out as the temperature raises.
This is a consequence of the fact that the Matsubara frequencies are given by $i \omega_j = i(2j+1) \pi T$.
The \texttt{EPW} code alternatively offers the possibility to enforce the number of frequency points using the variable \texttt{nswi} instead of \texttt{wscut}. This corresponds to using a different frequency cutoff for each temperature. We choose the former option in Fig.~\ref{aniso-MgB2}.

The superconducting density of states or the critical temperature have to be computed on the real energy axis. At the Fermi level we have that $\varepsilon_{n\mathbf{k}}=0$, and therefore Eq.~\eqref{gapequation} reduces to $\omega = E_{n\mathbf{k}} = \Re\Delta_{n\mathbf{k}}(\omega)$.
This implies that one has to look at the change of the superconducting gap with temperature at such frequencies. 
However, as shown on the bottom of Fig.~\ref{aniso-MgB2}, the dashed line $\omega = \Re\Delta_{n\mathbf{k}}(\omega)$ is steep, and the superconducting gap is almost constant in that energy range, with a value almost equal to its value at $\omega=0$. 
In the limit of $\omega=0$, the superconducting gap along the imaginary and real axes have the same value. 
As a result, one can use the superconducting gap $\Delta_{n\mathbf{k}}(i\omega=0)$ along the imaginary axis, which is computationally more affordable, to deduce the critical temperature without having to perform analytic continuation to the real axis.

\begin{figure}[b!]
  \centering
  \includegraphics[width=0.99\linewidth]{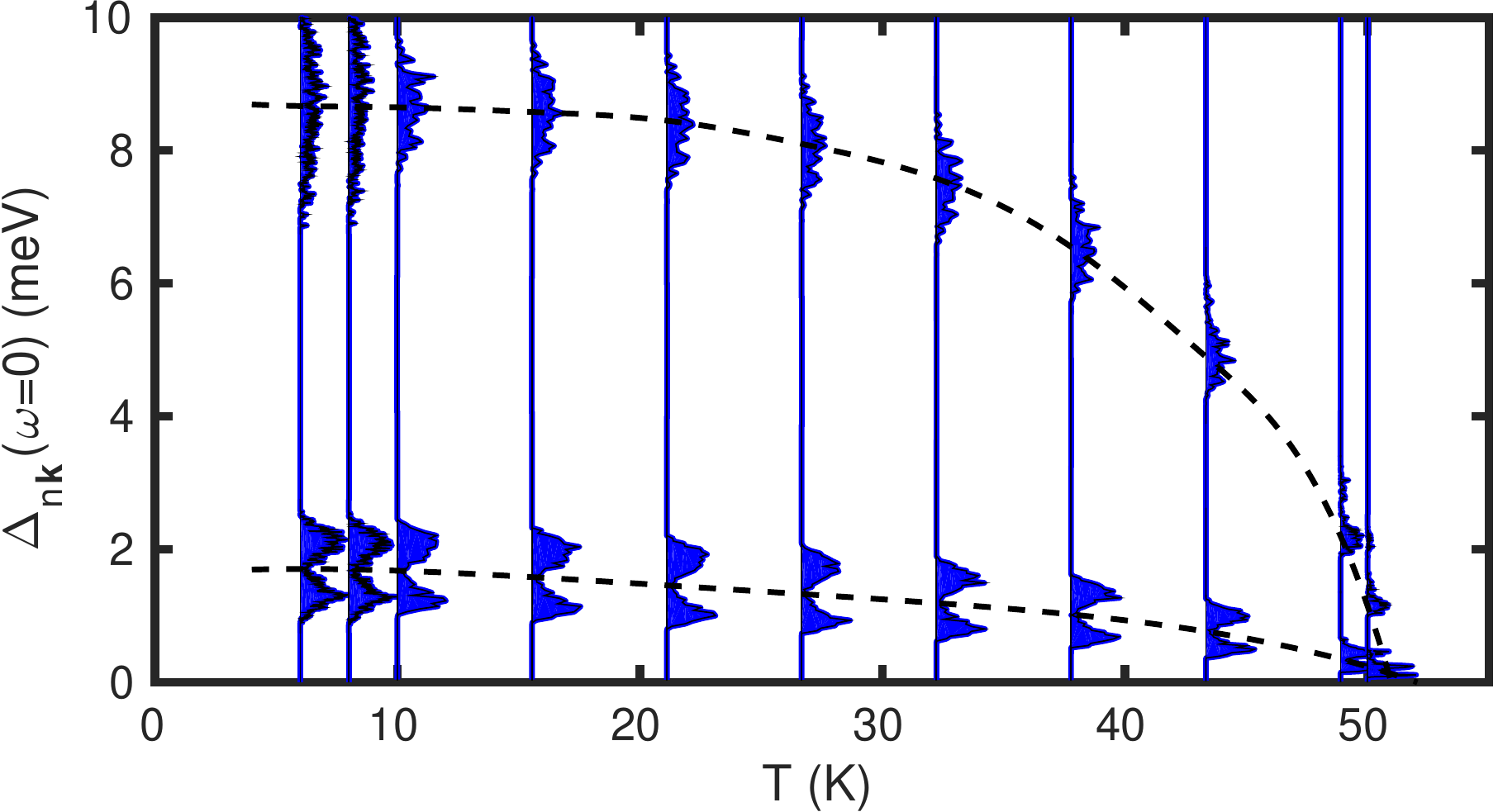}
  \caption{\label{gaps-MgB2} (Color online) Calculated distribution of the anisotropic superconducting gap $\Delta_{n\mathbf{k}}(\omega=0)$ of MgB$_2$ on the Fermi surface as a function of temperature. The superconducting gaps are obtained from analytic continuation using Pad\'e functions, and the Coulomb potential has been set to $\mu_{\rm c}^*=0.16$. The dashed lines are guides to the eye.}
\end{figure}

Another way to look at these data is presented in Fig.~\ref{gaps-MgB2}. Here we show the energy distribution of the calculated superconducting gap $\Delta_{n\mathbf{k}}(\omega=0)$ of MgB$_2$ as a function of temperature. The superconducting gaps are obtained from analytic continuation using Pad\'e functions, and the Coulomb potential has been set to $\mu_{\rm c}^*=0.16$ following Ref.~\cite{Margine2013}. 
One can clearly see the two-gap nature of MgB$_2$.
The gaps vanish at the critical temperature $T_{\rm c}=51$~K. This value overestimates the experimental value of 39~K~\cite{Nagamatsu2001}, but it matches previous first-principles calculations based on the Migdal-Eliashberg~\cite{Choi2002,Margine2013} or SCDFT~\cite{Floris2005} formalism.
It has been proposed that the overestimation of the critical temperature could be due to phonon anharmonicity~\cite{Choi2002}, diagrams beyond the Migdal approximation~\cite{Cappelluti2002,Pickett2003,Saitta2008,Calandra2010}, or an anisotropic Coulomb parameter. 
In comparison, the semi-empirical Allen-Dynes formula of Eq.~\eqref{ADTc} yields a critical temperature of 15.8~K using an isotropic Coulomb parameter of $\mu_{\rm c}^*=0.16$.


\begin{figure}[t!]
  \centering
  \includegraphics[width=0.99\linewidth]{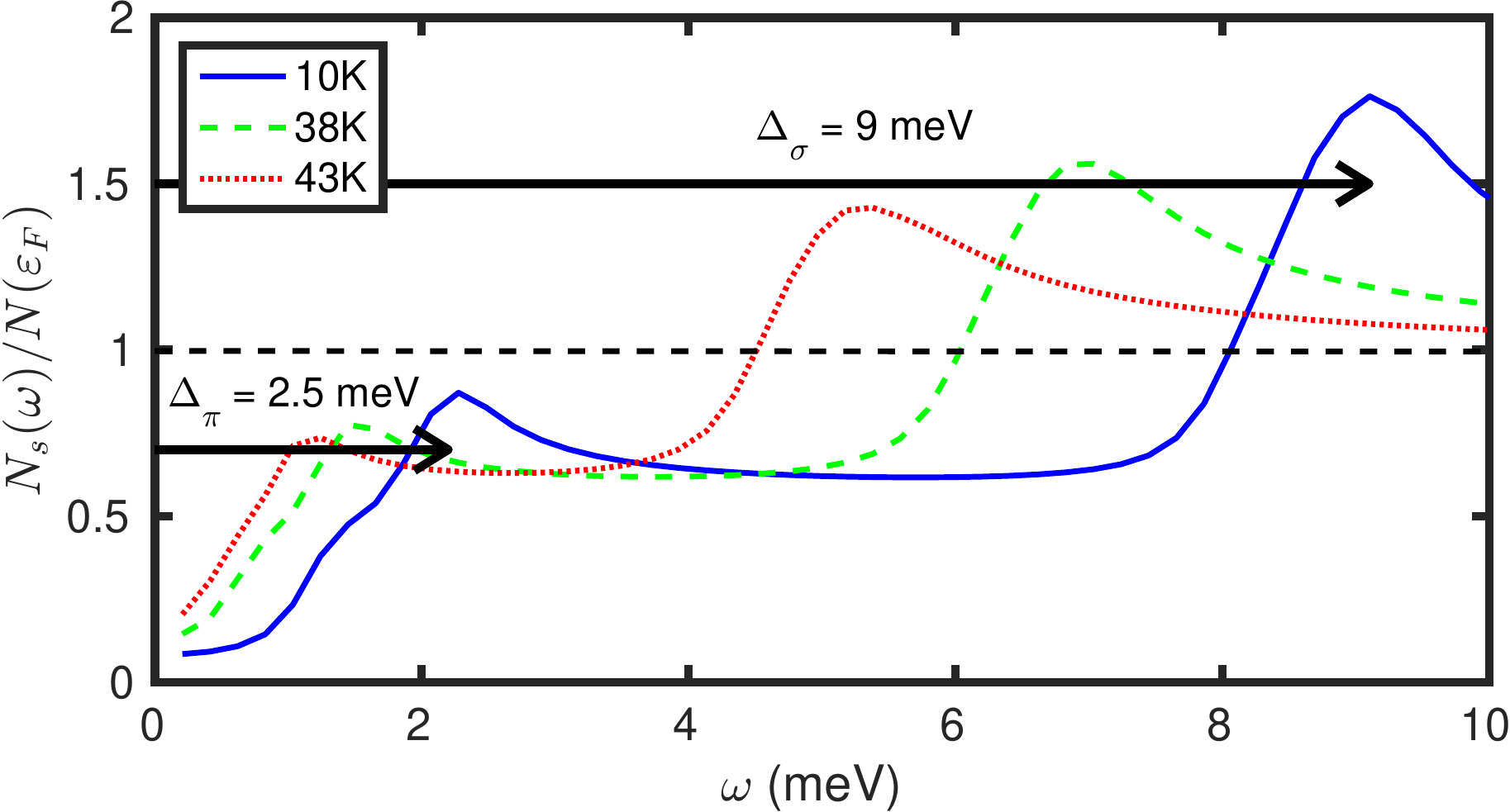}
  \caption{\label{tunnel-MgB2} (Color online) Quasiparticle superconducting DOS at different temperatures using Eq.~\eqref{tunnelingdensity2} with $\Delta(\omega=0)$. The dashed black line is the normal state DOS. The superconducting DOS $N_s(\omega)$ has been scaled so that its high energy tail coincides with the DOS in the normal state.}
\end{figure}

\begin{figure}[t!]
  \centering
  \includegraphics[width=0.99\linewidth]{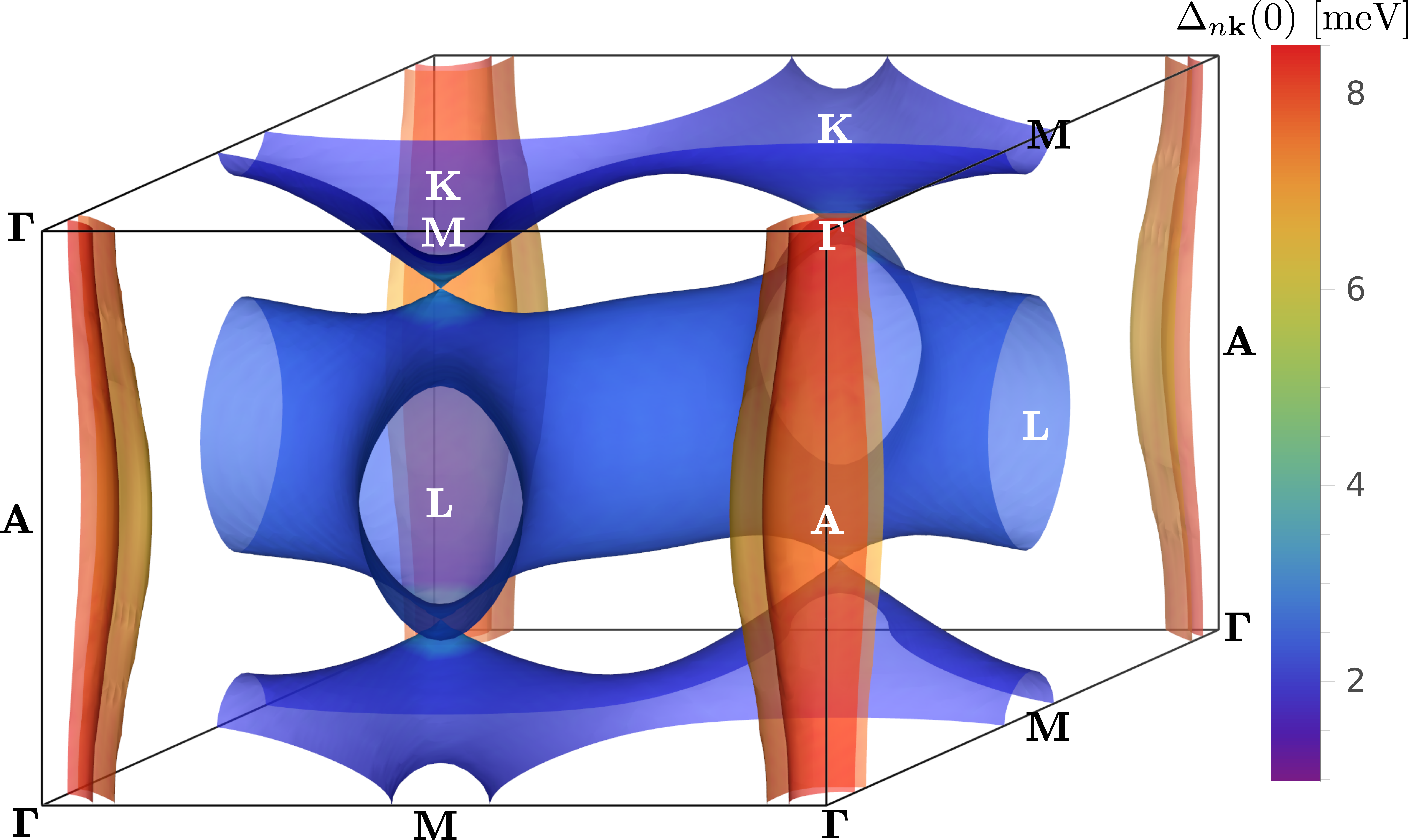}
  \caption{\label{FS-MgB2} (Color online) The superconducting energy gaps of MgB$_2$ (expressed in meV) on the Fermi surface for T = 10~K. The Fermi surface consists of two $\sigma$ sheets along the $\Gamma-\Gamma$ lines and two tubular structures defining a honeycomb lattice and arising from the $\pi$ states. The rendering has been done using the VESTA software~\cite{Momma2011}.}
\end{figure}

We can compute the DOS in the superconducting state at different temperatures using Eq.~\eqref{tunnelingdensity2} with $\Delta(\omega=0)$,  as shown in Fig.~\ref{tunnel-MgB2}. The DOS can be compared directly with experimental tunneling measurements~\cite{Giubileo2001}. 
The dashed black line is the normal state DOS. The superconducting DOS $N_s(\omega)$ has been scaled so that its high energy tail coincides with the DOS in the normal state.
In this figure we notice that, as temperature increases, the superconducting gap in the DOS tends to close.
The two-gap nature of MgB$_2$ can clearly be seen. The Fermi-surface averages of these gaps at 10~K are $\Delta_\pi$=2.5~meV and $\Delta_\sigma$=9~meV. 
This is close to the experimental gaps, measured at 4.2~K using Andreev reflection and scanning tunneling microscopy, that range from 2.3 to 2.8~meV for the $\pi$ band and 7.0 to 7.1~meV for the $\sigma$ band~\cite{Szabo2001,Iavarone2002,Gonnelli2002}.

The superconducting gap can alternatively be shown on the Fermi surface (see Fig.~\ref{FS-MgB2} for T = 10~K).
The Fermi surface consists of two $\sigma$ sheets along the $\Gamma-\Gamma$ line stemming from the $\sigma$-bond of the $p_{x,y}$ orbitals of boron, and two $\pi$ sheets along the K-M and H-L lines arising from the $p_z$ orbitals of boron. These results reproduce well previous studies~\cite{Choi2002,Choi2003}.

\begin{figure}[t!]
  \centering
  \includegraphics[width=0.99\linewidth]{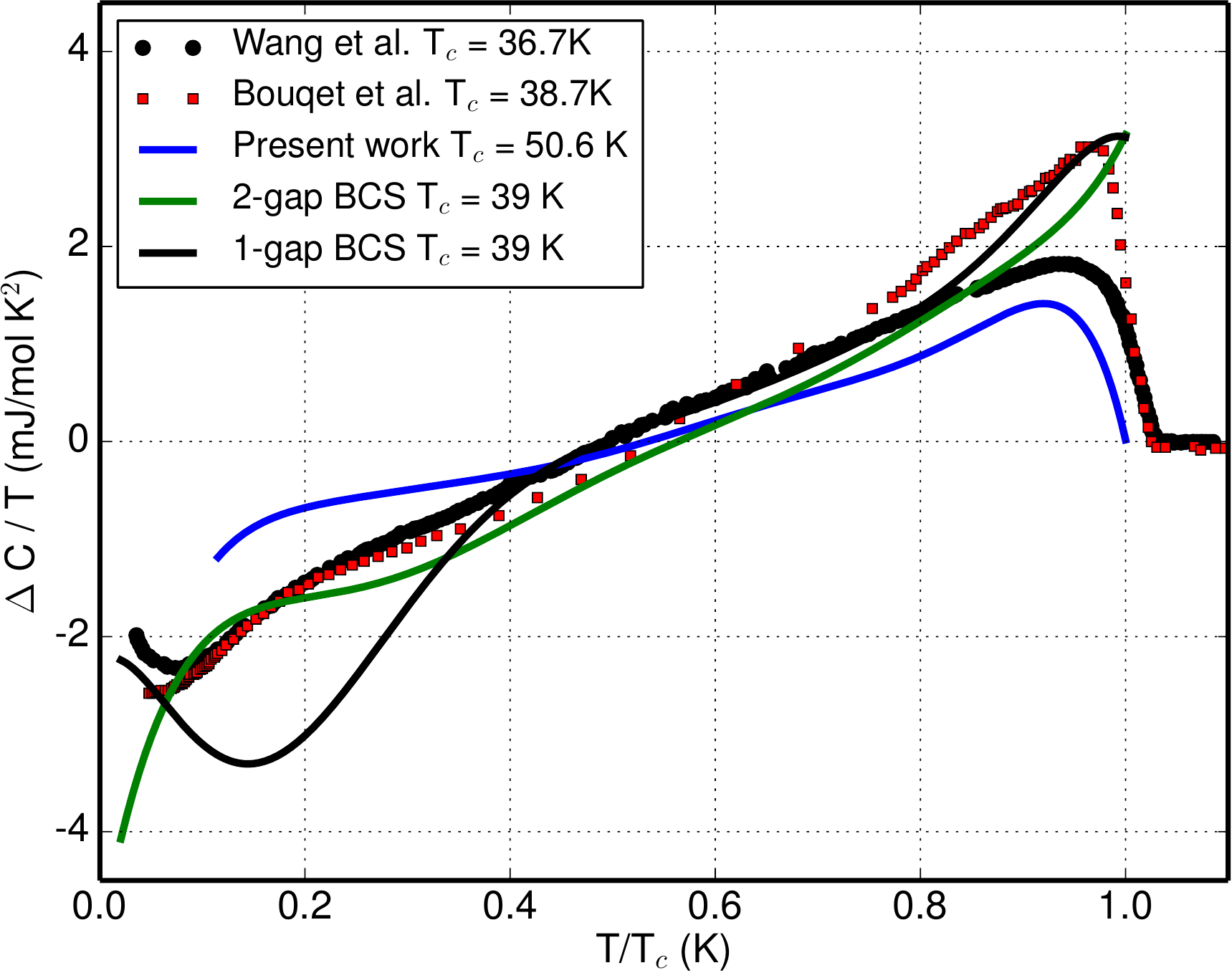}
  \caption{\label{specific-MgB2} (Color online) Calculated (straight blue line), 2-gap BCS model (straight green line), 1-gap BCS model (straight black line) and experimental (symbols) specific heat of MgB$_2$. The experimental measurements are from Refs.~\cite{Wang2001,Bouquet2001}. }
\end{figure}

Figure~\ref{specific-MgB2} shows the specific heat calculated using Eqs.~\eqref{free_energy} and \eqref{specific_heat} as well as the two BCS models. The specific heat was obtained by numerically computing the second-order derivative of the free energy as a post-processing step. Although our computed specific heat is shifted up by about 10~K, the anisotropic Eliashberg-Migdal theory reproduces well the experimental results of Refs.~\cite{Wang2001,Bouquet2001}. 
To compare with the previous work of Ref.~\cite{Choi2003}, 
we can compute the 1-gap BCS model using Eq.~\eqref{BCS1},
 $p=3.3$, T$_c$=39~K and $\Delta(T=0)=5$~meV. 
We can also construct a 2-gap BCS model using the values from Ref.~\cite{Choi2003} of T$_c$=39~K, $\Delta_1(T=0)=1.8$ and $\Delta_2(T=0)=6.8$ meV, $p_1=1.8$ and $p_2=2.9$. More information is given in the Appendix~\ref{Appendix2}.

Notably, both the 2-gap BCS model and our first-principle results recover the observed hump below 10~K whereas the 1-gap BCS model is clearly insufficient. 

\section{Conclusions}
\label{Conclusions}

In this article we presented the new capabilities offered by the \texttt{EPW} software in the field of electron-phonon calculations.
We focused on the new developments made in the code since the previous technical paper in 2010~\cite{Noffsinger2010}.  
The code is now fully integrated into \texttt{Quantum ESPRESSO}, while retaining its identity via its dedicated website and specific user support through its dedicated forum.

The code now supports spin-orbit coupling and time-reversal symmetry. Special attention was paid to code optimization and parallelization as well as code stability through the creation of a test farm. 

The three most notable new features of the code are the capability to compute anisotropic superconducting properties within the Migdal-Eliashberg theory, the Wannier interpolation in polar materials, and the calculations of electronic transport properties via the electronic velocities. 
Five physically relevant examples showcasing new features were presented in detail; four of them are available as tutorials in the \texttt{EPW} software distribution. 

Electronic mobilities using more refined theories like the Boltzmann transport equation or Kubo formalism, exact electronic velocities, first-principle evaluation of the Coulomb potential, $\mathbf{G}$-vector parallelization and parallelization for massively parallel architectures are planned for a future release.


\section{Acknowledgments}
\label{Acknowledgments}
We would like to thank P. Giannozzi and F. Spiga for help and support during the integration of \texttt{EPW} into \texttt{Quantum ESPRESSO}, and S. de Gironcoli, A. Dal Corso, W. Li, M. Zacharias and S. Baroni for fruitful discussions. 
We would like to thank M. Filip and M. Schlipf for generating the lead, gallium and nitrogen pseudopotentials used in this paper.

The research leading to these results has received funding from the Leverhulme Trust (Grant RL-2012-001), the UK Engineering and Physical
Sciences Research Council (grants No. EP/J009857/1 and EP/M020517/1), and the Graphene Flagship (EU FP7 grant no. 604391).
The authors acknowledge the use of the University of Oxford Advanced Research Computing (ARC) facility (http://dx.doi.org/
10.5281/zenodo.22558), the ARCHER UK National Supercomputing Service under the `AMSEC' Leadership project and the `CTOA' RAP project, and the Cartesius Dutch National Supercomputer under the PRACE DECI-13 project.

\section{Appendix: Time-reversal and separable pseudopotential in \texttt{EPW}}
\label{Appendix}

Building on the derivation proposed in Ref.~\cite{Giustino2007}, we extend the equations to treat the particular case of crystals with no inversion symmetry. In these cases, it is possible to take advantage of time-reversal symmetry to decrease the number of irreducible $\mathbf{q}$-points required on the coarse grid. 

Within linear response, the change of the self-consistent potential with respect to a collective ionic displacement $\Delta\boldsymbol{\tau}_{\kappa p}^{\mathbf{q}\nu}$ with phonon momentum $\mathbf{q}$ and mode index $\nu$ can be obtained from the screening of the bare ionic potential $V^{\text{ion}}_\kappa(\mathbf{r})$ as:
\begin{multline}
\partial_{\mathbf{q}\nu} V(\mathbf{r}) = \\
 \frac{\partial}{\partial \eta}  \int d\mathbf{r}' \varepsilon^{-1}(\mathbf{r},\mathbf{r'}) \sum_{\kappa,\mathbf{R}_p} V^{\text{ion}}_\kappa(\mathbf{r}'-\boldsymbol{\tau}_{\kappa p}-\eta \Delta \boldsymbol{\tau}_{\kappa p}^{\mathbf{q}\nu}),
\end{multline}
where $\varepsilon^{-1}(\mathbf{r},\mathbf{r'})$ is the electronic dielectric function and $\boldsymbol{\tau}_{\kappa p}$ is the coordinate of each ion $\kappa$ in the unit cell centered at $\mathbf{R}_p$.

The ionic displacement can be expressed in the basis of the periodic and normalized vibrational eigenmodes $\mathbf{e}_{\mathbf{q}\kappa}^\nu$:
\begin{equation}\label{iondispl}
\Delta \boldsymbol{\tau}_{\kappa p}^{\mathbf{q}\nu} =  \sqrt{\frac{m_0}{m_\kappa}}\mathbf{e}_{\mathbf{q}\kappa}^\nu e^{i\mathbf{q}\cdot \mathbf{R}_p}, 
\end{equation}
where $m_\kappa$ is the ionic mass of atom $\kappa$.

Following the convention defined in Ref.~\cite{MARADUDIN1968}, the application of a symmetry operation $\{\mathcal{S}|\mathbf{v}\}$ on a real space point $\mathbf{r}$ gives $\{\mathcal{S}|\mathbf{v}\}\mathbf{r} = \mathcal{S}\mathbf{r} + \mathbf{v}$, with $\mathcal{S}$ the rotation part and $\mathbf{v}$ the possible fractional translation. 

Considering that the symmetry operation $\{\mathcal{S}|\mathbf{v}\}$ maps $\kappa p$ into $\kappa' p'$, that the dielectric function is invariant under $\{\mathcal{S}|\mathbf{v}\}$, and that the ionic part of the pseudopotential is invariant under $\mathcal{S}$, we obtain:  
\begin{equation}
\partial_{\mathcal{S}\mathbf{q}\nu} V(\mathbf{r}) = \partial_{\mathbf{q}\nu} V(\{\mathcal{S}|\mathbf{v}\}^{-1}\mathbf{r}).
\end{equation}

The electron-phonon matrix element introduced in Eq.~\eqref{elphmatrix} for  momentum transfer $\mathcal{S}\mathbf{q}$ then reads:
\begin{multline}
	g_{mn,\nu}(\mathbf{k,\mathcal{S}q}) =  \\
	 \frac{1}{\sqrt{2 \omega_{\mathcal{S}\mathbf{q}\nu}} }\big\langle \psi_{m\mathbf{k+\mathcal{S}q}} (\mathbf{r}) \big| \partial_{\mathbf{q}\nu}V (\{\mathcal{S}|\mathbf{v}\}^{-1}\mathbf{r}) \big|  (\mathbf{r}) \psi_{n\mathbf{k}}\big\rangle, 
\end{multline}
which, after a change of spatial variable and noting that $\omega_{\mathcal{S}\mathbf{q}\nu}=\omega_{\mathbf{q}\nu}$, becomes:
\begin{multline}\label{basicGKK}
	g_{mn,\nu}(\mathbf{k,\mathcal{S}q}) = \\
	 \frac{1}{\sqrt{2\omega_{\mathbf{q}\nu}}} \big\langle  \psi_{m\mathbf{k+\mathcal{S}q}} (\{\mathcal{S}|\mathbf{v}\}\mathbf{r}) \big| \partial_{\mathbf{q}\nu}V (\mathbf{r}) \big| \psi_{n\mathbf{k}}(\{\mathcal{S}|\mathbf{v}\}\mathbf{r}) \big\rangle. 
\end{multline}

To simplify the above equation, in principle one could use the following relationship:
\begin{equation}\label{wfrelation}
\psi_{m\mathbf{k+\mathcal{S}q}}(\{\mathcal{S}|\mathbf{v}\}\mathbf{r}) = \psi_{m\mathbf{\mathcal{S}^{-1}k+q}}(\mathbf{r}). 
\end{equation}

However, such equality implies a phase relationship which is not enforced in electronic structure codes, and therefore does not hold in general. 

In case of a separable pseudopotential based on the Kleinman and Bylander (KB) procedure~\cite{Kleinman1982}, by using the completeness of the planewaves basis $\sum_{\mathbf{G}}|\mathbf{k}+\mathbf{G}\rangle \langle \mathbf{k}+\mathbf{G} |= 1$, Eq.~\eqref{basicGKK} becomes:
\begin{multline}
	g_{mn,\nu}(\mathbf{k,\mathcal{S}q}) = \\
	 \frac{1}{\sqrt{2\omega_{\mathbf{q}\nu}} } \Big[ \big\langle  \psi_{m\mathbf{k+\mathcal{S}q}}(\{\mathcal{S}|\mathbf{v}\}\mathbf{r}) \big| \partial_{\mathbf{q}\nu}V^{\text{scf}} (\mathbf{r}) \big| \psi_{n\mathbf{k}} (\{\mathcal{S}|\mathbf{v}\}\mathbf{r}) \big\rangle  \\	 
	+ \sum_{\mathbf{G}\mathbf{G'}}  
\big\langle  \psi_{m\mathbf{k+\mathcal{S}q}} (\{\mathcal{S} |\mathbf{v}\}\mathbf{r}) \big|  \mathbf{k}+ \mathcal{S}\mathbf{q} + \mathbf{G}(\{\mathcal{S} |\mathbf{v}\}\mathbf{r}) 	\big\rangle \\
	 \times \big\langle \mathbf{k}+ \mathcal{S}\mathbf{q} + \mathbf{G} (\{\mathcal{S} |\mathbf{v}\}\mathbf{r})  \big| \partial_{\mathbf{q}\nu}V^{\text{ion}} (\mathbf{r})  \big|    \mathbf{k}+\mathbf{G}' (\{\mathcal{S} |\mathbf{v}\}\mathbf{r})  \big\rangle \\
	    \times \big\langle   \mathbf{k} + \mathbf{G}'(\{\mathcal{S}|\mathbf{v}\}\mathbf{r})   \big|  \psi_{n\mathbf{k}}(\{\mathcal{S}|\mathbf{v}\}\mathbf{r})  \big\rangle  \Big],
\end{multline}
where $V^{\text{scf}} (\mathbf{r})$ is the screening contribution (local in DFT) that is computed in real space for convenience and then Fourier transformed back. Since the ionic part of the potential $V^{\text{ion}}$ needs to be computed at $\mathbf{r}$ we have introduced a complete plane wave basis set. The basis functions $|\mathbf{k}+\mathbf{G}\rangle$ respect the relationship given in Eq.~\eqref{wfrelation} and therefore we can deduce:
\begin{multline}\label{separableGKK}
	g_{mn,\nu}(\mathbf{k,\mathcal{S}q}) =\\
	  \frac{1}{\sqrt{2\omega_{\mathbf{q}\nu}}}
	 \Big[ \big\langle  \psi_{m\mathbf{k+\mathcal{S}q}}(\{\mathcal{S}|\mathbf{v}\}\mathbf{r}) \big| \partial_{\mathbf{q}\nu}V^{\text{scf}} (\mathbf{r}) \big| \psi_{n\mathbf{k}}(\{\mathcal{S}|\mathbf{v}\}\mathbf{r}) \big\rangle  \\	 
	+\sum_{\mathbf{G}\mathbf{G'}} 
	 \big\langle \psi_{m\mathbf{k+\mathcal{S}q}}(\{\mathcal{S} |\mathbf{v}\}\mathbf{r}) \big|   \mathbf{k}+ \mathcal{S}\mathbf{q} + \mathbf{G}(\{\mathcal{S} |\mathbf{v}\}\mathbf{r})  \big\rangle \\
	 \times \big\langle  \mathcal{S}^{-1}\mathbf{k}+\mathbf{q}+\mathbf{G} (\mathbf{r}) \big| \partial_{\mathbf{q}\nu}V^{\text{ion}} (\mathbf{r}) \big| \mathcal{S}^{-1}\mathbf{k}+\mathbf{G}' (\mathbf{r})  \big\rangle \\
	   \times \big\langle \mathbf{k} + \mathbf{G}'(\{\mathcal{S} |\mathbf{v}\}\mathbf{r})   \big| \psi_{n\mathbf{k}}  (\{\mathcal{S}|\mathbf{v}\}\mathbf{r})  \big\rangle \Big].
\end{multline}

The second term can be expressed in reciprocal space as follows, see Eq.~(A14) of Ref.~\cite{Baroni2001}:
\begin{multline}
\big\langle  \mathcal{S}^{-1}\mathbf{k}+\mathbf{q}+\mathbf{G}   \big|\partial_{\mathbf{q}\nu}V^{\text{ion}} \big| \mathcal{S}^{-1}\mathbf{k}+\mathbf{G}'  \big\rangle =\\
 \sum_\kappa -i \big( q_\nu + G_\nu -G_\nu' ) e^{-i(\mathbf{q+G-G'})\cdot \boldsymbol{\tau}_\kappa}
  \bigg[ \tilde{v}_\kappa^{\text{loc}}(\mathbf{q+G-G'}) \\
  + \sum_l \sum_{m=-l}^l \tilde{v}_{\kappa,lm} (\mathcal{S}^{-1}\mathbf{k}+\mathbf{q}+\mathbf{G},\mathcal{S}^{-1}\mathbf{k}+\mathbf{G}')   \bigg].
\end{multline}

Here the exponential is the structure factor, $\tilde{v}_\kappa^{\text{loc}}$ is the Fourier transform of the local part of the pseudopotential, and $\tilde{v}_{\kappa,lm}$ is the non-local part defined as:
\begin{multline}
  \tilde{v}_{\kappa,lm} (\mathcal{S}^{-1}\mathbf{k}+\mathbf{q}+\mathbf{G},\mathcal{S}^{-1}\mathbf{k}+\mathbf{G}') = \\
  c_{\kappa,lm}\, \tilde{\beta}_{\kappa,lm}^* (\mathcal{S}^{-1}\mathbf{k}+\mathbf{q}+\mathbf{G})\,\tilde{\beta}_{\kappa,lm} (\mathcal{S}^{-1}\mathbf{k}+\mathbf{G}'),
\end{multline}
where $\tilde{\beta}_{\kappa,lm}$ is the Fourier transform of the KB projectors:
\begin{equation}
\beta_{\kappa,lm}(\mathbf{r}) =  \hat{V}_{\kappa l}(r) \phi_{lm}^{\text{ps}}(\mathbf{r}),
\end{equation}
and the coefficient $c_{\kappa,lm}$ are given by:
\begin{equation}
c_{\kappa,lm} = \frac{1}{\big\langle \phi_{lm}^{\text{ps}}(\mathbf{r}) \big|  \hat{V}_{\kappa l}(r) \big| \phi_{lm}^{\text{ps}}(\mathbf{r}) \big\rangle }.
\end{equation}

In these expression $\phi_{lm}^{\text{ps}}(\mathbf{r})$ is the atomic pseudo-wavefunction, and $\hat{V}_{\kappa l}(r)$ is the angular momentum component of the pseudopotential. 

Having expressed the rotated matrix element in the KB form, we can write the time-reversed version of Eq.~\eqref{separableGKK} that we implemented in the code:
\begin{multline}\label{separableGKKreversal}
	g_{mn,\nu}(\mathbf{k,-\mathcal{S}q}) = \\
	 \frac{1}{\sqrt{2\omega_{\mathbf{q}\nu}}}\Big[ \big\langle \psi_{m\mathbf{k-\mathcal{S}q}}(\{\mathcal{S}|\mathbf{v}\}\mathbf{r}) \big| \Big(\partial_{\mathbf{q}\nu}V^{\text{scf}} (\mathbf{r}) \Big)^* \big|  \psi_{n\mathbf{k}}(\{\mathcal{S}|\mathbf{v}\}\mathbf{r}) \big\rangle  \\
	+  \sum_{\mathbf{G}\mathbf{G'}} 
	\big\langle  \psi_{m\mathbf{k-\mathcal{S}q}}(\{\mathcal{S} |\mathbf{v}\}\mathbf{r}) \big|  \mathbf{k}- \mathcal{S}\mathbf{q} +\mathbf{G}(\{\mathcal{S} |\mathbf{v}\}\mathbf{r})\big\rangle  \\
	 \times \big\langle \mathcal{S}^{-1}\mathbf{k}-\mathbf{q}+\mathbf{G}(\mathbf{r})  \big| \partial_{-\mathbf{q}\nu}V^{\text{ion}} (\mathbf{r}) \big|   \mathcal{S}^{-1}\mathbf{k}+\mathbf{G}'(\mathbf{r})   \big\rangle \\
	  \times \big\langle \mathbf{k} +\mathbf{G}' (\{\mathcal{S} |\mathbf{v}\}\mathbf{r})  \big| \psi_{n\mathbf{k}}  (\{\mathcal{S}|\mathbf{v}\}\mathbf{r})  \big\rangle  \Big].
\end{multline}

\section{Appendix: Specific heat in MgB$_2$ and the two-gap BCS model}
\label{Appendix2}

By free energy minimization, Bardeen, Cooper and Schrieffer obtained an integro-differential equation that links the inverse of the interaction strength $\lambda$ to the superconducting gap $\Delta(T=0)$ (see Eq.~(3.27) of Ref.~\cite{Bardeen1957}):
\begin{align}\label{BCS0}
\frac{1}{\lambda} &= \frac{1}{N(\varepsilon_F)\Omega_{\text{BZ}}} \nonumber \\
                  &= \int_0^{\hbar\omega}d\xi \frac{\tanh\Big[\frac{1}{2k_B T}[\xi^2+\Delta^2(T)]^{1/2}\Big]}{[\xi^2+\Delta^2(T)]^{1/2}},
\end{align}
where $\lambda$ is the isotropic integrated $\lambda_{\mathbf{q}\nu}$ from Eq.~\eqref{lambdaEQ}.
Here $\hbar$ and $k_B$ are indicated explicitly to be consistent with the seminal equation in Ref.~\cite{Bardeen1957}.

They also derived what is now known as the universal BCS relation for weak coupling (i.e. $\hbar\omega \gg \Delta(T=0)$), see Eq.~(3.30) of Ref.~\cite{Bardeen1957}: 
\begin{equation}\label{BCS1}
\Delta(T=0) = 1.75 k_B T_c.
\end{equation}

Finally, using a variational approximation to the true ground-state, they deduced (see Eq.~(2.40) of Ref.~\cite{Bardeen1957}):
\begin{equation}\label{BCS2}
\frac{\hbar\omega}{\Delta(T=0)} = \sinh\Big[\frac{1}{\lambda}\Big].
\end{equation}

We can also define the following two dimensionless variables:
\begin{equation}
\delta = \frac{\Delta(T)}{\Delta(T=0)} \quad \text{ and } \tau = \frac{T}{T_c}.
\end{equation}

Inserting Eqs.~\eqref{BCS1} and \eqref{BCS2} inside Eq.~\eqref{BCS0} and performing a change of variable $z = \frac{\xi}{\Delta(T)}$ leads to~\cite{Johnston2013}
\begin{equation}\label{BCS4}
\frac{1}{\lambda} = \int_0^{\delta^{-1}\sinh(1/\lambda)} \frac{dz}{(1+z^2)^{1/2}} \tanh\Big[\frac{1.75}{2}\frac{\delta}{\tau} (1+z^2)^{1/2} \Big].
\end{equation}

\begin{figure}[t!]
  \centering
  \includegraphics[width=0.99\linewidth]{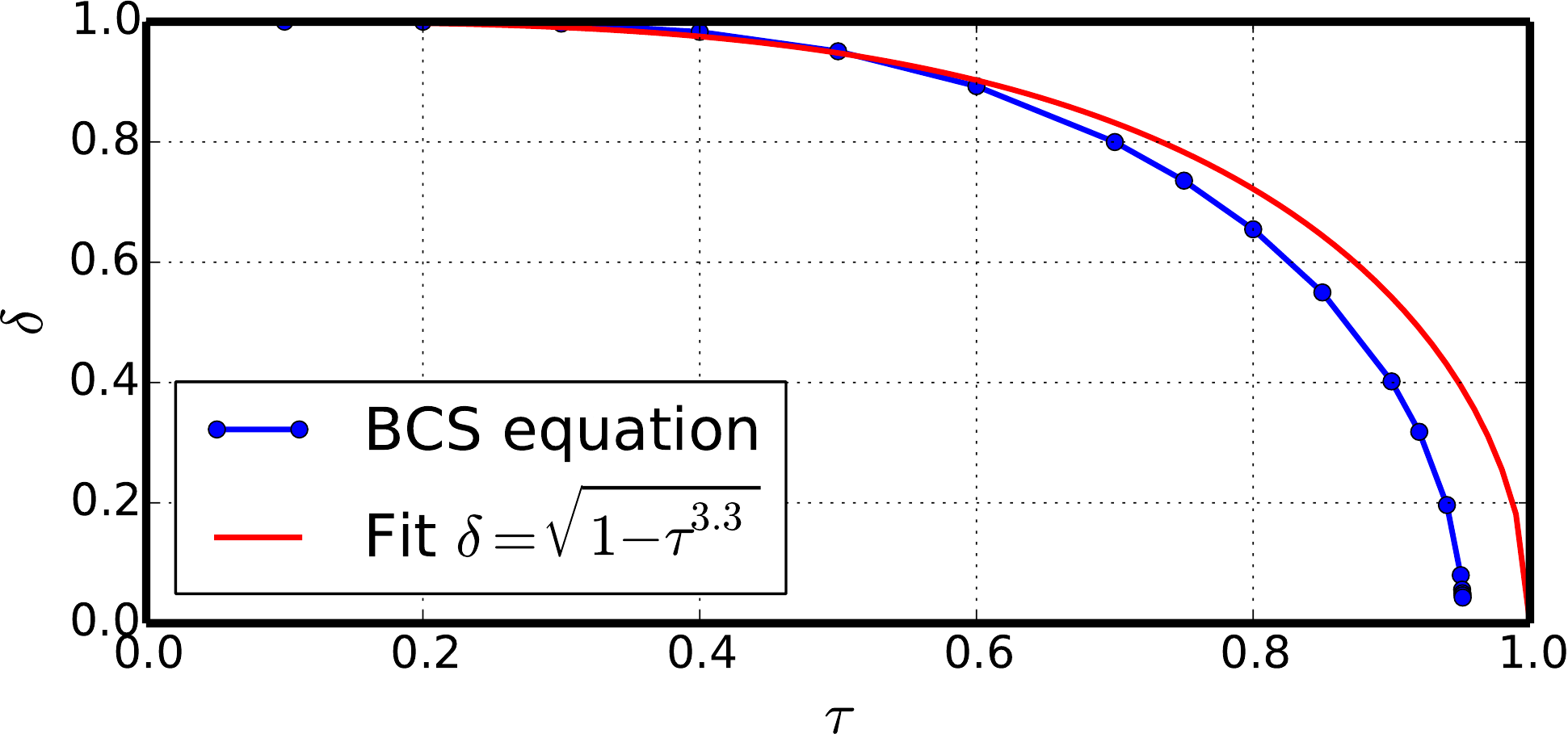}
  \caption{\label{MgB2-BCS-fit} (Color online) Adimentional BCS Calculated (blue line) and fitted (red line) superconducting gap ratio $\delta = \frac{\Delta(T)}{\Delta(T=0)}$ with respect to superconducting temperature ratio $\tau = \frac{T}{T_c}$ of MgB$_2$. }
\end{figure}

In the case of MgB$_2$, $\lambda = 0.75$ and therefore Eq.~\eqref{BCS4} can be solved by fixing $\tau$ and solving for $\delta$. 
However, Eq.~\eqref{BCS4} is expected to work for weak coupling superconductors in the range of $\lambda \le 0.3$ (determined empirically, see Ref.~\cite{Gennes1999}, p.112 for more details).
Indeed, as can be seen on Fig.~\ref{MgB2-BCS-fit}, the equation has no solution for $\tau$ close to 1. 

An alternative fitting formula can be used~\cite{Choi2003}:
\begin{equation}\label{deltaTeq}
\frac{\Delta(T)}{\Delta(T=0)} =\sqrt{1-\tau^p},
\end{equation}
where $p$ has to be fitted. By fitting on the solution of Eq.~\eqref{BCS4} in the low $\tau$ value range, we found $p=3.3$ (see Fig.~\eqref{MgB2-BCS-fit}) for $\lambda=0.75$.

Once we have a value for the isotropic superconducting gap $\Delta(T)$, Eqs.~\eqref{coupledAnisoEqs1} and \eqref{free_energy} can be simplified to:
\begin{multline}
\Delta F = -\pi T \sum_j \Big[ \sqrt{\omega_j^2+\Delta^2(T)}-\omega_j \Big]\\
\times \Big[ Z_j - Z_j^N \frac{\omega_j}{\sqrt{\omega_j^2+\Delta^2(T)}} \Big],
\end{multline}
with
\begin{equation}
Z_j = 1+\frac{\pi T}{\omega_j} \sum_{j'} \frac{\omega_{j'}}{\sqrt{\omega_{j'}^2+\Delta^2(T)}}\lambda
\end{equation}
and
\begin{equation}
Z_j^N = 1 + \frac{\pi T}{\omega_j}\lambda,
\end{equation}
where the sum over $j$ and $j'$ has to be numerically truncated so that
\begin{equation}
\omega_j = (2j+1)\pi T \le \omega_c.
\end{equation}
In our case we used a truncation energy of $\omega_c = 0.3$~eV.

\bibliographystyle{elsarticle-num} 
\bibliography{article_v16}


\end{document}